\documentclass[12pt]{iopart}
\usepackage{iopams}
\usepackage{amsfonts}
\usepackage{tikz}
\usepackage{graphicx}
\usepackage{caption}
\usepackage{pgf,tikz}
\usepackage{float}
\usepackage{color}
\usepackage{hyperref}
\usepackage{lineno}
\usepackage{cite}
\usetikzlibrary{arrows}
  
\begin{document}
\title[]{Relativistic sonic geometry for isothermal accretion in the Schwarzschild metric}
\author{Md Arif Shaikh$^{1,2}$, Ivleena Firdousi$^3$ and 
Tapas Kumar Das$^{1,2}$}

\address{$^1$ Harish-Chandra Research Institute, Chhatnag Road, Allahabad 211019, India}
\address{$^2$ Homi Bhabha National Institute, Training School Complex, Anushaktinagar, Mumbai 400094, India}
\address{$^3$ Savitribai Phule Pune University, Pune 411007, India}
\ead{\mailto{arifshaikh@hri.res.in}, \mailto{ivleena.firdousi@gmail.com} and \mailto{tapas@hri.res.in}}

\begin{abstract}
In this work, we perform linear perturbation on general relativistic isothermal accretion
onto a non-rotating astrophysical black hole to study the salient features of the corresponding
emergent acoustic metric. For spherically symmetric accretion as well as for the axially
symmetric matter flow for three different geometric configuration of matter, we perturb the
velocity potential, the mass accretion rate, and the integral solution of the time independent
part of the general relativistic Euler equation to obtain such acoustic geometry. We provide
the procedure to locate the acoustic horizon and identify such horizon with the transonic
surfaces of the accreting matter through the construction of the corresponding causal structures.
We then discuss how one can compute the value of the acoustic surface gravity in terms of the
accretion variable corresponding to the background flow solutions - i.e., stationary integral
transonic accretion solutions for different matter geometries. We show that the salient features of
the acoustic geometry is independent of the physical variable we perturb, but sensitively depends on
the geometric configuration of the black hole accretion disc.
\end{abstract}


\vspace{2pc}

\submitto{\CQG}
%\maketitle

\section{Introduction}
Linear perturbation of inhomogeneous barotropic, inviscid irrotational fluid produces acoustic geometry embedded within such transonic fluid. The propagation of perturbation can be described by a space-time metric known as the acoustic metric. Such metric appears to be identical (apart from a conformal factor) to the Painlev{\'e}-Gullstrand\cite{Painleve,Gullstrand,Lemaitre,Israel1987,Kraus} representation of the Schwarzschild metric \cite{Schwarzschild} and possesses a horizon. Such horizons are known as acoustic horizon and may emit the sound quanta phonon. The spectra of such phonon radiation is somewhat similar to the radiation from black holes as proposed by Hawking \cite{Hawking1974,Hawking1975}. Such transonic fluid system with acoustic geometry embedded within it is analogous to the gravitational system where the gravity like phenomena emerges through the intrinsic curvature of the space-time metric. Such systems mimic the black hole spacetime in many ways and are known as analogue systems.

The analogue gravity phenomena was first discovered by Moncrief \cite{Moncrief1980} and Unruh \cite{Unruh}, and later represented using a more formal set up by Visser \cite{Visser1998} for fluid flow in the Newtonian (flat) background. Bili\'{c} \cite{Bilic1999} modified such formulation for flow in the general relativistic regime. A comprehensive discussion of the development of the theory of analogue gravity are available in \cite{Barcelo,Analogue-gravity-phenomenology,Unruh-Schutzhold,Novello-visser}. For an excellent non technical article on this topic, see \cite{echo-black-hole}. Majority of the works on analogue gravity, however, have been performed for physical systems not directly subjected to the gravitational field and the gravity like effects in such systems appears as an emergent phenomena. Off late, it was understood that the study of the analogue effect for accretion of matter onto astrophysical black holes may be of significant interest since the accreting black hole system itself contains a gravitational horizon and the transonic accretion onto black hole will give rise to additional sonic horizons. Accreting black holes are thus the only physical systems found in the universe where both kind of horizons, the gravitational one and the acoustic one, may be simultaneously present in the same physical system. accreting black hole system may be helpful in comparing the properties of these  two kind of horizons since the same accreting material encounters both of them. Study of analogue gravity phenomena for accreting black holes may thus be of significant  importance. In the following section, we shall briefly outline the procedure we follow to analyze the accreting black hole systems as a natural example of classical (non-quantum) analogue system.

\section{Accreting black holes as analogue systems}\label{Accreting black holes as analogue systems}
The inner boundary conditions imposed by the gravitational horizon of astrophysical black holes imply (\cite{Frank1985accretion,Liang1980transonic}) that accreting material crosses the horizon supersonically. Except for the special cases for which the accreting material originates from the supersonic stellar wind \cite{lamers1999introduction,accretion-japan2008}, black hole accretion is thus necessarily be transonic\cite{Liang1980transonic}. Linear perturbation of such transonic accretion may produce the acoustic spacetime embedded within the accreting fluid which will possess its acoustic horizon at the location where the transonic surface is formed for steady state flow. The essential idea in connection to this phenomena is the following: 

For hydrodynamic accretion flow, one can write down the Euler and the continuity equation to describe the dynamics of such accretion. It is assumed that the steady state has been achieved so that the stationary integral accretion solutions can provide the basic transonic features of accretion flow. Such background steady state solutions are then linearly perturbed to obtain the corresponding acoustic geometry. While perturbing the flow, certain physical quantity is to be identified which is being perturbed to accomplish the task. We shall discuss this issue in somewhat detail in the subsequent sections.

Once the acoustic geometry is studied, the location of the horizon is then identified, and it can easily be shown that such acoustic horizons are the transonic surfaces, where, for the background steady flow, the accreting material makes a transition from the subsonic to the supersonic state. Number and nature of such transonic surfaces depends on the symmetry of the problem and on the geometric configuration of the infalling matter. The simplest possible case is referred to the spherically symmetric accretion within Newtonian construct. Such flow was first described by Bondi \cite{Bondi1952} and is usually referred as Bondi flow or Bondi accretion. The general relativistic version of Bondi type flow was discussed by Michel \cite{Michel1972} and hence is referred by Michel flow in the literature. Michel Flow is considered to be the simplest possible general relativistic accretion configuration onto compact astrophysical objects. For such flow, the general relativistic Euler and the continuity equations are constructed from the energy momentum tensor of a perfect fluid described by certain barotropic equation of state - either adiabatic or isothermal. The flow is assumed to be a steady one and the integral solution of the time independent parts of the relativistic Euler and the continuity equation provides the stationary configuration of the steady transonic flow and the corresponding phase portrait. See, e.g.,\cite{TKD2001Spherical-wind}.

From the knowledge of such stationary solutions, one can calculate the acoustic surface gravity of such flow and can study the dependence of location of the transonic surfaces on various accretion parameters, see, e.g., \cite{Das-2004}. It is to be mentioned that as of now, Michel flow has been studied only for the adiabatic equation of state and isothermal accretion with spherical symmetry has not been discussed in the literature. In our present work, we study the Michel flow of isothermal fluid, perturb the background steady state solutions for such flow to understand the emergent acoustic geometry, and to construct the expression of the corresponding acoustic surface gravity from the dynamical concept - i.e, through linear perturbation of the steady state background solutions of Michel flow. The first part of our work concentrates to accomplish such task.

Unlike Bondi or Michel flow, axially symmetric accretion, however can have more than one transonic surfaces, see, e.g.,\cite{TKD2002,TKD-Mitra-2003,TKD2012-Czerny} for detailed discussion from the astrophysical context. For such accretion, the number of acoustic horizons may be more than one - up to three in principle, two black hole horizons at the inner and the outer sonic points, and an acoustic white hole horizon at the shock location. the stationary integral solutions corresponding to such multi-transonic flow with shock thus provides the embedded acoustic geometry with more than one sonic horizon. Once again, the isothermal general relativistic flow has not been studied in this context and in the present work we exactly do so. We linear perturb the background isothermal steady state solutions for axially symmetric accretion and obtain the corresponding acoustic geometry with multiple values of the acoustic surface gravity associated with various acoustic horizons as mentioned above. This we do for relativistic flow in the Schwarzschild metric.

For axially symmetric flow, the geometric configuration of the matter can vary on certain assumptions used to describe the flow. Below we briefly describe three different geometric configuration of matter as usually considered in the literature (\cite{Nag_role_of_flow_geometry} and references therein for further details)
\begin{enumerate}
\item \textit{Flow with constant thickness:} In this model of accretion disc the thickness of the disc is assumed to be constant. In other words the height of the disc does not change with the radial distance measured from the horizon along the equatorial plane. This kind of geometry resembles a right circular cylinder of a fixed height with the axis of symmetry along the $z$-axis. This is the simplest possible flow geometry one can choose to describe accretion disc in astrophysics. As one can see, the height of the disc, being constant, does not change when we linear perturb different flow variables in the linear perturbation analysis we perform in the later sections. If we denote the local height of the disc (half of the thickness of the disc), as measured from the equatorial plane along the vertical direction, by $H$ then for this model $H=\rm{constant}$.
\item \textit{Wedge shaped flow with conical geometry:} In the Bondi flow \cite{Bondi1952} or Michel flow \cite{Michel1972}, the flow geometry is spherically symmetric and the introduction of angular momentum destroys the spherical symmetry of the flow making the flow axially symmetric. However the flow can still be assumed to be quasi-spherical in the sense that though it deviates from the absolute spherical symmetry, the ratio of the height of the disc and the radial distance remains constant. In other words the local flow height $H$ is given by $H\propto r$, where $r$ is the radial distance. The proportionality constant is obtained from the solid angle subtended by the flow at the center. In this case also the height does not depend on accretion variables such as velocity and density. Therefore here also the height remains unaffected while we do the linear perturbation analysis.
\item \textit{Flow in hydrostatic equilibrium along vertical direction:} This model of flow geometry is the most complicated one. In this model the flow is assumed to be in hydrostatic equilibrium along vertical direction. The flow thickness is therefore a function of radial distance as well other accretion variables as discussed further in Sec \ref{Sec:Definition-mass-accretion-rate}.
\end{enumerate}
Acoustic spacetime corresponding to the aforementioned three different geometric configuration has been discussed for accretion under the influence of the Post Newtonian pseudo-Schwarzschild \cite{Nag_role_of_flow_geometry} as well as pseudo-Kerr \cite{Sonali2016} black hole potentials. Such works were performed within Newtonian framework. In their recent works, {Ananda} \etal \cite{deepika-sph2015} have shown how the perturbation of mass accretion rate leads to  acoustic geometry for general relativistic astrophysical accretion of polytropic matter onto non rotating black holes. Datta \etal \cite{Satadal-arif-2016} considered the linear perturbation of the integral solution of the time independent Bernoulli's equation to study the acoustic geometry corresponding to the adiabatic accretion, for both Newtonian, as well as for background Schwarzschild metric. Motivated by the aforementioned literature, what we intend to present in this work is an overall comprehensive treatment of the generation of curved acoustic manifold embedded within general relativistic isothermal accretion onto a non rotating black hole for all possible geometric configuration of axially symmetric flow structure. Not only that, such acoustic spacetime has been obtained by linear perturbing all relevant physical quantities. In usual literature, the velocity potential corresponding to the irrotational flow is normally perturbed to obtain the corresponding wave equation and the acoustic metric. In the present work, we perturb, separately, three quantities - the velocity potential, the mass accretion rate, and the integral solution of the time component of relativistic Euler equation, which following \cite{Anderson1989}, we designate as the relativistic Bernoulli's constant in this work. Such constant is a first integral of motion for the steady state accretion flow. The other first integral of motion is the mass accretion rate. We show that the intrinsic acoustic geometry remains invariant (apart from a conformal factor) for a particular type of symmetry and the background black hole metric, and does not depend on which of the three quantities, viz, velocity potential, mass accretion rate or the relativistic Bernoulli's constant is being linear perturbed to obtain the embedded acoustic geometry. 

The location and the number of the acoustic horizon are found to depend, quite sensitively, on the geometric configuration of the matter flow. Consequently, the value of the acoustic surface gravity also gets influenced by the matter geometry. Such thing is thus found to be a characteristic feature of the axially symmetric accretion only and no such matter geometry dependence of of the acoustic surface gravity is obtained for spherically symmetric flow. Hawking radiation does not depend on the medium and therefore it is said to be universal. However analogue effects depends on the medium through the dispersion relation. Such deviation has been studied for fluid systems in flat Minkowski spacetime for very simplified dispersion relations, see, e.g., \cite{Leonhardt2012-Robertson,Robertson2012}. However study of such deviation in black hole accretion analogue system is way more involved as in this system not only the background spacetime geometry is curved but also the matter flow can have different geometry which influences the acoustic surface gravity. Thus study of acoustic surface gravity for different flow geometry for isothermal accretion is important to understand such deviation of universality of analogue Hawking radiation in curved background spacetime.

%..................................................
\section{Summary of the methodology used}\label{Sec:methodology}
Starting from the expression of the energy momentum tensor of the ideal fluid, we first derive the relativistic Euler and the continuity equation in the schwarzschild metric corresponding to the spherical and axially symmetric flow of isothermal fluid on astrophysical black holes. We then linear perturb the velocity potential, the mass accretion rate and the relativistic Bernoulli's constant  to obtain the corresponding acoustic metric. We then find the location of acoustic horizon and study the causal structure around the horizon. Finally, we calculate the corresponding acoustic surface gravity in terms of the Schwarzschild metric and the unperturbed values of various accretion variables. By unperturbed values we mean the values corresponding to the time independent part of the the variables.

Energy preserving flow is not the only kind of astrophysical accretion perceived. The other form of barotropic flows - the isothermal accretion finds its importance in large scale astrophysical dynamics where temperature instead of the total specific energy remains invariant for the steady flow. To fulfill the inviscid constant angular momentum criteria we will consider low angular momentum accretion.
The first theory of standard accretion disc was given in 1973 \cite{Sakura-Sunyaev-1973}. However a realistic model of viscous accretion with dissipation is still not fully understood. Moreover such practically inviscid accretion may be observed at our galactic center\cite{Monika}. Inclusion of viscosity is expected to affect the location of the acoustic horizon. This is due to the fact that viscosity helps to transport angular momentum outwards which in turn lowers angular momentum in the inner region of the disc. However due to large radial velocity near the horizon the infall time scale is much smaller than the viscous time scale. Even at large distances the radial velocity is large due to relatively low rotational energy \cite{Beloborodov,Igumenshchev,Proga}. Hence our assumption of inviscid accretion flow is not unjustified from the astrophysical point of view.   Motivated by these facts, in the present work we study the emergence of black hole like acoustic metric in irrotational inviscid isothermal accretion in Schwarzschild background. 

In what follows, we discuss basic properties of axially symmetric black hole accretion and introduce the required physical variables or parameters which will be used in the subsequent sections.

We will concentrate on the stationary flow (background fluid) configuration on the equatorial plane of the disc. The radial distance measured from the black hole event horizon (gravitational) will be scaled by $GM_{\rm{BH}}/c^2$ where $G$ is the universal gravitational constant, $M_{\rm{BH}}$ is the mass of the black hole, and $c$ is the velocity of light in vacuum. All relevant velocities will be scaled by $c$, and $G=c=M_{\rm{BH}}=1$ will be adopted. Analytical treatment of the stationary integral solutions for more than one variable is intractable. We hence deal with an `effective' two dimensional solutions (In literature which is called 1.5 dimensional solutions, see \cite{TKD2002} and references therein). In such scheme, the dynamical variables are defined on the equatorial plane only and hence only radial velocity $u_0$ (as defined in Eq. (\ref{u-in-co-rot-frame})) and the radial sound speed $c_s$ (as defined  in Eq. (\ref{clayperon-eq})) are considered. We thus define the radial Mach number $\mathcal{M}$ as the ratio of the radial velocity and the radial sound speed, that is $\mathcal{M}=\frac{u_0}{c_s}$. Hereafter by Mach number we mean radial Mach number only. The transonic flow structure thus be described in terms of radial Mach number only. Following similar arguments, we consider the radial acoustic perturbation and any non-radial perturbation will be assumed to be non-contributing. The vertical structure of the disc, however, is considered through the averaging process. The thermodynamic quantities are vertically averaged over the flow thickness. Symbolically speaking, the vertically averaged density is
\begin{equation}
\bar{\rho} = \frac{\int_{-\frac{H}{2}}^{\frac{H}{2}} \rho dh}{\int_{-\frac{H}{2}}^{\frac{H}{2}} dh}
\end{equation}
where $H$ is the thickness of the flow. See, e.g., \cite{matsumoto1984viscous,gammie_and_popham} for further details of the vertical averaging procedures in axially symmetric black hole accretion.

Hereafter $\rho,\epsilon,p,T$ will be considered as the matter density, mass energy density, pressure, and the bulk ion temperature of the flow, respectively and the accreting matter will be considered as a single temperature non-self gravitating isothermal fluid. The isothermal sound speed $c_s$ may be expressed in terms of the bulk ion temperature through the Clapeyron-Mendeleev equation \cite{bazarov1964thermodynamics,gibbs2014elementary}
\begin{equation}\label{clayperon-eq}
c_s = \sqrt{\frac{k_B}{\mu m_H}T}
\end{equation}
where $k_B$ is the Boltzmann constant, $m_H$ is the mass of the hydrogen atom ($m_H\approx m_p$ for one temperature fluid) and $\mu$ is the mean molecular weight. The space invariance of the temperature for the steady state accretion implies that the isothermal sound speed is position independent for the background flow.

We consider the background black hole spacetime to be stationary and spherically symmetric. However the accretion may be spherical or axially symmetric if it have non zero angular momentum and is endowed with two commuting Killing vector fields. The energy momentum tensor of an ideal fluid can be constructed as
\begin{equation}\label{energy-mom}
 T^{\mu\nu}=(p+\epsilon)v^\mu v^\nu+pg^{\mu\nu}
\end{equation}
$g^{\mu\nu}$ are the metric elements corresponding to the background black hole spacetime.
We consider the following general form of the background static space-time
\begin{equation}\label{metric}
ds^2=-g_{tt}dt^2+g_{rr}dr^2+g_{\theta\theta}d\theta^2+g_{\phi\phi}d\phi^2
\end{equation}
where the metric elements are functions of $r$ and $\theta$. For Schwarzschild metric the metric elements are given by
\begin{eqnarray}\label{metric_elements}
g_{tt}=\frac{1}{g_{rr}}=\left(1-\frac{2}{r}\right)\\
g_{\theta\theta} = \frac{g_{\phi\phi}}{\sin^2\theta}=r^2.
\end{eqnarray}
On the equatorial plane ($\theta=\frac{\pi}{2}$), $g_{\phi\phi}=r^2$. $v^{\mu}$ is the timelike fibre bundle defined on the background manifold constructed by the family of (fluid) streamlines; in other words $v^{\mu}$ are the velocity vector fields which follows the normalization condition as $v^\mu v_{\mu} = -1$. The four-velocity components are given as $(v^t,v^r,v^\theta,v^\phi)$.  The specific angular momentum $\lambda$ of the axially symmetric flow is described as $\lambda = -\frac{v_\phi}{v_t}$ where $v_t,v_\phi$ are the temporal and azimuthal components of the covariant velocity field $v_\mu = (v_t,v_r,v_\theta,v_\phi)$. As already explained that we will concentrate on low angular momentum inviscid flow and hence $\lambda$ will be assumed to be constant for our background steady flow and will be parameterized by a suitable value corresponding to the sub-Keplerian accretion. The angular velocity thus can be defined as 
\begin{equation}\label{angular-velocity}
\Omega = \frac{v^\phi}{v^t}=\frac{g_{tt}}{g_{\phi\phi}}\lambda
\end{equation}
The advective velocity $u$ of the accreting fluid is related to the three velocity $v$ as
\begin{equation}\label{u-in-co-rot-frame}
u = \frac{v}{\sqrt{1-\lambda \Omega}}
\end{equation}
and the stationary value of the advective velocity will be denoted as $u_0$.
Hence the radial component of $v^\mu$ that is $v^r$ is related to $u$ as
\begin{equation}\label{v-r-in-co-rot-frame}
v^r= \frac{u}{\sqrt{g_{rr}(1-u^2)}}
\end{equation}
The advective velocity $u$ is the radial velocity of the fluid which is measured in a frame co-rotating with the accreting matter, see, e.g., \cite{gammie_and_popham} for further details. Using Eq. (\ref{angular-velocity}) and (\ref{v-r-in-co-rot-frame}) and the normalization condition $v^\mu v_\mu = -1$ gives
\begin{equation}\label{v-t-in-co-rot-frame}
v^t = \frac{1}{\sqrt{g_{tt}(1-u^2)(1-\lambda\Omega)}}
\end{equation}
which when used in Eq. (\ref{angular-velocity}) gives
\begin{equation}\label{v-phi-in-co-rot-frame}
v^\phi = \Omega v^t = \frac{\Omega}{\sqrt{g_{tt}(1-u^2)(1-\lambda\Omega)}}
\end{equation}
using these transformations we are able to describe the accretion flow completely in terms of $u$ and $\lambda$ (or $\Omega$).
\section{Basic equations}
The continuity equation ensuring the conservation of mass is given by 
\begin{equation}\label{cont eq}
\nabla_\mu(\rho v^\mu)=0
\end{equation}
The energy-momentum conservation equation is given by 
\begin{equation}\label{mom-cons-eq}
\nabla_\mu T^{\mu\nu}=0
\end{equation}
Substitution of Eq. (\ref{energy-mom}) in Eq. (\ref{mom-cons-eq}) provides
 the general relativistic Euler equation as 
\begin{equation}\label{eq 5}
(p+\epsilon)v^\mu \nabla_\mu v^\nu+(g^{\mu\nu}+v^\mu v^\nu) \nabla_\mu p=0
\end{equation}
The enthalpy is defined as 
\begin{equation}
h=\frac{(p+\epsilon)}{\rho}
\end{equation}
where $\rho$ is the fluid density. The isothermal sound speed is given by \cite{Yuan1996}
\begin{equation}\label{sound speed}
c_s^2=\frac{1}{h}\frac{\partial p}{\partial \rho}
\end{equation}
The relativistic Euler equation for isothermal fluid can thus be written as 
\begin{equation}\label{Eq:euler-isothermal}
v^\mu \nabla_\mu v^\nu+\frac{c_s^2}{\rho}(v^\mu v^\nu+g^{\mu\nu})\partial_\mu \rho=0
\end{equation}
which can be further written using the expression for covariant derivative {$\nabla_\mu v^\nu=\partial_\mu v^\nu+\Gamma ^\nu_{\mu \lambda}v^\lambda$} as 
\begin{equation}\label{rel euler eq}
v^\mu \partial_\mu v^\nu+\Gamma^\nu_{\mu\lambda} v^\mu v^\lambda+\frac{c_s^2}{\rho}(v^\mu v^\nu+g^{\mu\nu})\partial_\mu \rho=0
\end{equation}

\section{Velocity potential, mass accretion rate and the relativistic Bernoulli's constant}\label{three-quantities} 
In this section we shall find out the expressions of the velocity potential, the relativistic Bernoulli's constant and the mass accretion rate.  We shall show how these quantities are obtained from the irrotationality condition, the general relativistic Euler equation and the continuity equation respectively.

\subsection{Velocity potential}\label{Sec:Definition-velocity-potential}
In Newtonian framework vorticity $\vec{\omega}$ is defined as the curl of the velocity vector $\vec{v}$, i.e., $\vec{\omega}=\nabla\times \vec{v}$ \cite{Landau_fluid}. Thus it represents the local rotation of the fluid elements in the flow.
In general relativity the vorticity  is defined as \cite{Bilic1999,Liang2013}
\begin{equation}\label{vorticity}
\omega_{\mu\nu}=l^\rho_{~\mu} l^\sigma_{~\nu} v_{[\rho;\sigma]}
\end{equation}
where $v_{\mu;\nu}=\nabla_\nu v_{\mu}$ and  $v_{[\mu;\nu]}\equiv \frac{1}{2}[v_{\mu;\nu}-v_{\nu;\mu}]$ and  $l^\mu_{~\nu}$ is the projection operator which projects an arbitrary vector in space-time into its components in the subspace orthogonal to $v^\nu$ and it is given by $l^\mu_{~\nu}=\delta^\mu_{~\nu}+v^\mu v_\nu.$ Vorticity geometrically measures the twisting of the
congruence\cite{Iyer}. In general a flow may have non vanishing vorticity $\omega_{\mu\nu}$. A flow is said to be  irrotational if it has vanishing vorticity, that is $\omega_{\mu\nu}=0$. This gives the irrotationality condition to be (see \ref{Appendix:irrotationality-condition})
\begin{equation}\label{irrotationality}
 \partial_\mu( v_\nu \rho^{c_s^2})-\partial_\nu( v_\mu \rho^{c_s^2})=0
 \end{equation} 
 where we have used the fact that for isothermal fluid $c_s^2$ is  constant since the flow temperature remains invariant throughout. The above equation helps us to introduce a potential field which, in analogy to the non relativistic case, we call velocity potential $\psi$. Thus the velocity potential $\psi$ is defined by the relation 
\begin{equation}\label{definition-velocity-potential}
 v_\mu \rho^{c_s^2}=\partial_\mu \psi.
 \end{equation} 

\subsection{Relativistic Bernoulli's constant}\label{Sec:Definition-bernoulli-constant}
First we shall discuss it in the context of the spherically symmetric  Michel Flow. For Michel Flow due to spherical symmetry $v^\phi = v^\theta = 0$. The temporal component of the Eq. (\ref{rel euler eq}) obtained by using $\nu=t$ can be written as
\begin{equation}
v^r\partial_r v^t+v^t\partial_t v^t+\Gamma^t_{\mu\lambda} v^\mu v^\lambda+\frac{c_s^2}{\rho}[(v^t)^2-g^{tt}]\partial_t \rho+\frac{c_s^2}{\rho}v^rv^t\partial_r\rho=0
\end{equation}
The relevant Christoffel symbol is $\Gamma^t_{rt}=\frac{1}{2}g^{tt}\partial_t g_{tt}$. Also the normalization condition gives $g_{tt}(v^t)^2-1=g_{rr}(v^r)^2$. Thus the above equation can be rearranged as
\begin{equation}\label{time comp of ber eq sph}
v^t\partial_t v^t+\frac{c_s^2}{\rho}\frac{g_{rr}(v^r)^2}{g_{tt}}\partial_t\rho+v^rv^t\partial_r\{\ln (v_t\rho^{c_s^2})\}=0
\end{equation}
 For stationary flow, where derivatives with respect to $t$ vanish, the above equation gives the relativistic Bernoulli's constant
\begin{equation}
v_{t0}\rho_{0}^{c_s^2}=\rm{Constant}= \xi_0
\end{equation}
 where $v_{t0}$ and $\rho_0$ are the stationary values of $v_t$ and $\rho$ respectively. It should be noted that $\xi_0$ can not be identified with the actual specific energy which is not constant for isothermal flow.

For the axially symmetric flow we assume that there is no convection current along non-equatorial direction and hence $v^\theta$ is vanishing but $v^\phi$ is non zero. Thus using the normalization condition and the relevant Christofell symbol $\Gamma^t_{rt}=\frac{1}{2}g^{tt}\partial_r g_{tt}$ the temporal component of the the relativistic Euler equation for axially symmetric flow can be written as 
\begin{equation}\label{time com euler axi}
v^t\partial_t v^t+\frac{c_s^2}{\rho}\frac{\{g_{rr}(v^r)^2+g_{\phi\phi}(v^\phi)^2\}}{g_{tt}}\partial_t\rho+v^rv^t\partial_r\left\{\ln (v_t\rho^{c_s^2})\right\}=0
\end{equation}
Therefore for stationary axially symmetric flow also, as in Michel flow, $\xi_0=v_{t0} \rho_0^{c_s^2}$ is constant.

\subsection{mass accretion rate}\label{Sec:Definition-mass-accretion-rate}
The continuity equation given by Eq. (\ref{cont eq}) can be written as
\begin{equation}\label{cov-cont-eq}
 \frac{1}{\sqrt{-{g}}}\partial_\mu (\sqrt{-g}\rho v^\mu)=0
 \end{equation} 
 where $g = $ det$ g_{\mu\nu}$ = $-r^4\sin^2\theta$. For spherically symmetric accretion $\partial_\theta$ and $\partial_\phi$ terms do not contribute and hence the continuity equation becomes
\begin{equation}\label{cont eq in sph}
\frac{1}{\sqrt{-g}}\partial_t \left(\rho v^t \sqrt{-g} \right)+\frac{1}{\sqrt{-g}}\partial_r \left(\rho v^r\sqrt{-g} \right)=0.
\end{equation}
One can also multiply Eq. (\ref{cov-cont-eq}) by the co-variant volume element $\sqrt{-g}d^4x$ to get
\begin{equation}\label{co-variant-cont-int}
\partial_\mu (\sqrt{-g}\rho v^\mu)d^4x=0
\end{equation}
For stationary flow along with $\partial_\theta,\partial_\phi$ also the $\partial_t$ term vanish which leads to the equation 
\begin{equation}
\partial_r(\sqrt{-g}\rho_0 v_0^r)drd\theta d\phi=0
\end{equation}
writing $g = \tilde{g}\sin^2\theta$, where $\tilde{g}=-r^4$, gives $\partial_r(\sqrt{-\tilde{g}}\rho_0v_0^r \sin\theta)drd\theta d\phi=0$. We can integrate out the $\theta,\phi$ part over appropriate range which will give a purely geometrical factor $\tilde{\Omega}$ (tilde is used in order to not confuse it with the angular velocity $\Omega$ defined in Eq. (\ref{angular-velocity})). Thus we will have
\begin{equation}
\tilde{\Omega}\sqrt{-\tilde{g}}\rho_0 v_0^r=-\dot{M} 
\end{equation}
$\dot{M}$ represents the mas accretion rate, i.e., the mass flux per unit time for ingoing accretion solution. Note that the presence of a `dot' does not mean time derivative (since we are considering stationary solutions only), it simply signifies that that the amount of mass falling in through a certain surface remains invariant per unit time for steady state solutions. The negative sign implies the in fall of matter . As $\tilde{\Omega}$ is merely a geometrical factor we can absorb it in the right hand side to redefine the mass accretion rate to be $\Psi_0=\sqrt{-\tilde{g}}v_0^r\rho_0$ without any loss of generality.
 
For axially symmetric accretion, under the assumption that $v^\theta=0$ and the axial symmetry,  equation(\ref{cov-cont-eq}) can be written as 
\begin{equation}\label{conti-pre-avg}
 \partial_t(\rho v^t \sqrt{{-g}})+\partial_r(\rho v^r\sqrt{-{g}})=0.
 \end{equation}
 We now perform a vertical averaging following \cite{gammie_and_popham} by integrating the above equation over $\theta$. This averaging over a variable $f(r,\theta)$ is approximated as
 \begin{equation}
 \int f(r,\theta) d\theta  \approx H_\theta f(r,\theta = \frac{\pi}{2})
 \end{equation}
Where $H_\theta$ is characteristic angular scale of the flow and it is a function of local flow height $H$ and hence depends on the model of accretion we are working with. In other words $H_\theta$ contains information about the flow structure along the vertical direction. Thus vertical averaging provides us a way to work fully in the equatorial plane by containing information about vertical structure inside $H_\theta$. $H_\theta$ may also be thought as the appropriate weight function to get the correct flux of infalling matter while integrating the continuity equation. Thus after the vertical averaging the continuity Eq. (\ref{conti-pre-avg}) becomes
 \begin{equation}\label{cont eq in axi}
 \partial_t(\rho v^t \sqrt{-\tilde{g}}H_\theta)+\partial_r(\rho v^r\sqrt{-\tilde{g}} H_\theta)=0
 \end{equation}
For stationary flow it gives
\begin{equation}
\partial_r(\rho_0 v_0^r\sqrt{-\tilde{g}} (H_\theta)_0)=0,
\end{equation}
or
\begin{equation}
\rho_0 v_0^r\sqrt{-\tilde{g}}(H_\theta)_0=\rm{constant}.
\end{equation}
where $(H_\theta)_0$ represents the stationary value of $H_\theta$. To get the rate of infall of matter we also need to integrate over the azimuthal angle which introduces a geometrical factor that can be absorbed without any loss of generality to define the mass accretion rate as
\begin{equation}
\rho_0 v_0^r\sqrt{-\tilde{g}}(H_\theta)_0=\rm{constant}\equiv\Psi_0.
\end{equation}
As $H_\theta$ is the characteristic angular scale of the flow, for conical flow it is constant, whereas for flow with constant height $H_\theta \propto \frac{1}{r}$. The expression of $H_\theta$ for flow with vertical equilibrium is rather complicated. This is due to the fact that in order to maintain hydrostatic equilibrium along vertical direction the pressure gradient must be balanced by the component of gravitational force due to the accretor along the opposite direction. In Newtonian background one can use the Euler equation and find the relation between $H_\theta$ and other accretion variables. For general relativistic case it is quite involved and has been explicitly derived in \cite{Abramowicz1996ap} for Kerr background. Taking the Schwarzschild limit the relation is given by
\begin{equation}\label{H-theta}
 H_\theta^2v_\phi^2f(r)=\frac{p}{\rho}
\end{equation}
where $f(r)$ is independent of fluid variables. Therefore, due to the dependence of $H_{\theta}$ on $p,\rho$ and $v^\phi$, when we do linear perturbation analysis, described in the next section, it will also be perturbed as given in Eq. (\ref{H1}). For more details on $H_\theta$ for different flow geometry we refer to \cite{deepika_ax_schwarzchild}. 

\section{The acoustic metric from linear perturbation analysis}
In this section we summarize the results obtained by linear perturbation analysis while providing the detailed procedure of the perturbation analysis for the spherically symmetric Michel flow and axially symmetric flow in the \ref{Linear-perturbation-michel} and \ref{Liner-perturbation-axial} respectively. In both cases the linear perturbation analysis is done for three different quantities which we introduced in previous section (Sec. \ref{three-quantities}). The quantities we defined are velocity potential $\psi$, the relativistic Bernoulli's constant $\xi_0$ and the mass accretion rate $\Psi_0$. Let us write the velocity potential about its stationary value $\psi_0(r)$ as
\begin{equation}
\psi(r,t) = \psi_0(r)+\psi_1(r,t)
\end{equation}
where $\psi_1(r,t)$ represents small perturbation about the stationary value. Similarly $\xi_0$ and $\Psi_0$ could be considered as the stationary value of the quantity $\xi(r,t)$ and $\Psi(r,t)$ given by
\begin{equation}\label{Define-xi}
 \xi(r,t) = v_{t}(r,t)\rho(r,t)^{c_s^2} = \xi_0 + \xi_1(r,t)
 \end{equation} 
 and
 \begin{equation}\label{Define-mass}
 \Psi(r,t) = v^r(r,t)\rho(r,t)\sqrt{-\tilde{g}}=\Psi_0 + \Psi_1(r,t)
 \end{equation}
where $\xi_1$ and $\Psi_1$ are small perturbation about the stationary background value $\xi_0$ and $\Psi_0$. We also write $\rho$ and $v^\mu$ about their background value as 
\begin{equation}
\rho(r,t) = \rho_0(r) + \rho_1(r,t)
\end{equation}
and 
\begin{equation}
v^{\mu} = v^\mu_0 + v^\mu_1
\end{equation}
Using these equations in the irrotationality condition, the continuity equation and the relativistic Bernoulli's equation enables us to obtain the wave equation describing the propagation of the perturbations $\psi_1$, $\xi_1$ and $\Psi_1$. From the analysis it is seen that the wave equation is given by an equation of the general form
\begin{equation}\label{acoustic-wave-eq}
\partial_\mu( f^{\mu\nu}\partial_\nu \tilde{\varphi})=0
\end{equation}
where $\tilde{\varphi}$ may be $\psi_1,\xi_1$ or $\Psi_1$ and $\mu,\nu$ takes values $(t,r)$. On the other hand the propagation of a massless scalar field $\varphi$ in curved spacetime described by metric $g_{\mu\nu}$ is given by
\begin{equation}\label{massless-scalar-eq}
\frac{1}{\sqrt{-g}}\partial_\nu (\sqrt{-g}g^{\mu\nu}\partial_\nu \varphi)=0.
\end{equation}
Comparing these two equations Eq. (\ref{acoustic-wave-eq}) and Eq. (\ref{massless-scalar-eq}) one obtains the   acoustic metric $G^{\mu\nu}$ as
\begin{equation}\label{g and f rel}
 \sqrt{-G}G^{\mu\nu} = f^{\mu\nu}
 \end{equation} 
 where $G$ is determinant of the metric $G_{\mu\nu}$. Therefore our main aim is to find out these $f^{\mu\nu}$ as the acoustic metric $G^{\mu\nu}$ is related to $f^{\mu\nu}$ by just a conformal factor. To find the exact expression for $G^{\mu\nu}$ we need to know $G$. In general this should be accomplished by taking determinant of both side of the Eq. (\ref{g and f rel}). 
 However due to the fact that $f^{\mu\nu}$ is represented by a $2\times2$ matrix one can not determine $G$ from Eq. (\ref{g and f rel}). This can be understood more clearly if we take a case where $f^{\mu\nu}$ and $G^{\mu\nu}$ may be represented by a $n\times n$ matrix. In this case taking determinant of both sides of Eq. (\ref{g and f rel}) gives 
\begin{equation}\label{G}
-G = (-det f^{\mu\nu})^{\frac{2}{n-2}}
\end{equation}
which clearly encounters a problem when $n=2$. However there is no problem if $f^{\mu\nu}$ is (1+3) dimensional i.e $n=4$. This indicates, as mentioned in \cite{Barcelo}, that the problem of determining $G$ for $(1+1)$ dimensional $f^{\mu\nu}$ is not a fundamental one as $f^{\mu\nu}$ is (1+1) dimensional due the fact that we considered the accretion flow to have some symmetry as well as the background black hole spacetime to be spherically symmetric. These symmetry reduces the dimension of $f^{\mu\nu}$ form $(1+3)$ to $(1+1)$ though the actual background spacetime where the accretion takes place is $(1+3)$ dimensional. This is evident from  the calculation in Sec. \ref{Linear-pert-velo-michel} where $f^{\mu\nu}$, when no symmetry was assumed, is $(1+3)$ dimensional. As the background spacetime is $(1+3)$ dimensional one can always augment the (1+1)-dimensional geometry by two extra flat dimensions (see \ref{Appendix:metric-augmentation}) and hence use Eq. (\ref{G}) to get the acoustic metric \cite{Barcelo}.

In principal one, however, do not really need to find $G$ exactly to study conformally invariant features of acoustic metric as $G$ contributes only as a conformal factor to the acoustic metric $G^{\mu\nu}$. The location of the acoustic event horizon, the causal structure of the acoustic spacetime and the acoustic surface gravity are independent of the conformal factor. Therefore to study these features of the acoustic spacetime, which we do in the subsequent sections, one can ignore the conformal factor and work with the relevant part of $G_{\mu\nu}$ or $G^{\mu\nu}$. Hence we will neglect any conformal factor and work with only the relevant part of the acoustic metric. 

From the linear perturbation analysis in \ref{Linear-perturbation-michel} and \ref{Liner-perturbation-axial} we obtain the $f^{\mu\nu}$ for different cases. In the following we will give the expressions of the $f^{\mu\nu}$ and  the relevant $G^{\mu\nu}$ and $G_{\mu\nu}$ (where any conformal factor is neglected.) obtained from these $f^{\mu\nu}$.

\subsection{Michel flow}
In \ref{Linear-pert-velo-michel} we show that without considering any symmetry of the background spacetime or the accreting flow one obtains the expression for $f^{\mu\nu}$ by linear perturbing velocity potential as
\begin{equation}\label{f for velo}
f^{\mu\nu}=-\frac{(\sqrt{-g})}{(\rho^{c_s^2-1}c_s^2)}\left[-c_s^2g^{\mu\nu}+\left(1-c_s^2\right)v^\mu_0v^\nu_0\right]
\end{equation}
However when we consider the background black hole spacetime to be spherically symmetric and the accretion also to be spherically symmetric, the $f^{\mu\nu}$ reduces to $(1+1)$ dimension. The linear perturbation of the relativistic Bernoulli's constant in \ref{Linear-pert-bern-michel} and the mass accretion rate in \ref{Linear-pert-mass_michel} also gives $(1+1)$ dimensional $f^{\mu\nu}$. These $f^{\mu\nu}$ obtained by linear perturbing three different quantities differs only by an overall factor which we denote as $k_i(r)$ where $i=1,2,3$ represents the cases of velocity potential, relativistic Bernoulli's constant and the mass accretion rate respectively. Thus the $f^{\mu\nu}$ obtained in these three cases can be given as
\begin{equation}
f^{\mu\nu}=k_i(r)\left[\begin{array}{ccc}
\frac{c_s^2+g_{tt}(v_0^t)^2(1-c_s^2)}{g_{tt}} & v^r_0v_0^t(1-c_s^2)\\
v^r_0v_0^t(1-c_s^2) & \frac{-c_s^2+g_{rr}(v^r_0)^2(1-c_s^2)}{g_{rr}}
\end{array}\right]
\end{equation}
where $k_1(r)=k_2(r)=-(\sqrt{-g})/(c_s^2\rho^{c_s^2-1})$ and $k_3(r)=-(g_{rr}v^r_0)/(v_0^t)$. The fact that $k_1=k_2$ but not equal to $k_3$ is discussed in Sec. \ref{Concluding-Remarks}. This overall factors $k_i$ contributes to the final acoustic metric only as conformal factors and hence not important to study conformally invariant properties of the acoustic metric and therefore the $f^{\mu\nu}$ obtained in the three different cases leads to same features of the acoustic spacetime. Hence neglecting these $k_i$ and the $G$, as both of these factors contribute to conformal factors only, we can get the acoustic metric $G^{\mu\nu}$ using Eq. (\ref{g and f rel}) as
\begin{equation}\label{G-upper-michel}
G^{\mu\nu}=-\left[\begin{array}{cc}
\frac{c_s^2+g_{tt}(v_0^t)^2(1-c_s^2)}{g_{tt}} & v_0^rv_0^t(1-c_s^2)\\
v_0^rv_0^t(1-c_s^2) & \frac{-c_s^2+g_{rr}(v_0^r)^2(1-c_s^2)}{g_{rr}}
\end{array}\right]
\end{equation}
and therefore $G_{\mu\nu}$ (where we again neglect any overall factor arising in the process of taking inverse of $G^{\mu\nu}$) as
\begin{equation}\label{metric_sph}
G_{\mu\nu}=\left[\begin{array}{cc}
\frac{-c_s^2+g_{rr}(v_0^r)^2(1-c_s^2)}{g_{rr}} & -v_0^rv_0^t(1-c_s^2)\\
-v_0^rv_0^t(1-c_s^2) &  \frac{c_s^2+g_{tt}(v_0^t)^2(1-c_s^2)}{g_{tt}}
\end{array}\right]
\end{equation}
For the completeness we also give the $G^{\mu\nu}$ and $G_{\mu\nu}$ obtained using $f^{\mu\nu}$ in Eq. (\ref{f for velo}) (where as usual any conformal factor is neglected) as
\begin{equation}
G^{\mu\nu}=g^{\mu\nu}+(1-\frac{1}{c_s^2})v^\mu v^\nu
\end{equation}
and 
\begin{equation}
G_{\mu\nu}=g_{\mu\nu}+(1-c_s^2)v_\mu v_\nu
\end{equation}
\subsection{Axially symmetric flow}
For axially symmetric accretion we consider the configuration geometry of the flow also. Therefore the $f^{\mu\nu}$ contains a term $\beta$ (as defined in Eq. (\ref{H1}) ) which may be different for different flow configuration and hence represents the dependence of $f^{\mu\nu}$ on the flow configuration. As shown in \ref{Liner-perturbation-axial}, $\beta = 0$ for both flow with constant thickness and wedge shaped conical flow and thus these two configurations have exactly same $f^{\mu\nu}$. However for flow with hydrostatic equilibrium along vertical direction $\beta = c_s^2$ and thus $f^{\mu\nu}$ differs from other two cases. Hence the final acoustic metric obtained from $f^{\mu\nu}$ is dependent on the flow configuration. $f^{\mu\nu}$ for the linear perturbation of velocity potential (\ref{Linear-velo-axi}), relativistic Bernoulli's constant (\ref{Linear-bern-axi}) and the mass accretion rate (\ref{Linear-mass-axi}) are given as
\begin{equation}\label{velo f axi}
f^{\mu\nu}=\tilde{k}_i(r)\left[\begin{array}{cc}
-g^{tt}+(1-\frac{1+\beta}{c_s^2})(v_0^t)^2 & (1-\frac{1+\beta}{c_s^2})v^r_0v^t_0 \\
(1-\frac{1+\beta}{c_s^2})v^r_0v^t_0 & g^{rr}+(1-\frac{1+\beta}{c_s^2})(v^r_0)^2
\end{array}\right]
\end{equation} 
where $i=1,2,3$ for the linear perturbation of velocity potential, relativistic Bernoulli's constant and the mass accretion rate respectively and $\tilde{k}_1(r)=\sqrt{-\tilde{g}}/\rho_0^{c_s^2-1}$, $\tilde{k}_2(r)=(\sqrt{-\tilde{g}}(H_\theta)_0)/(\rho_0^{c_s^2-1})$,  $\tilde{k}_3(r)=(g_{rr}v^r_0c_s^2)/(v_0^t\Lambda)$. ($\Lambda$ is given in \ref{Linear-mass-axi}). From $f^{\mu\nu}$ the acoustic metric $G^{\mu\nu}$ and ${G_{\mu\nu}}$ are obtained as
\begin{equation}\label{G-upper-axi}
G^{\mu\nu}=\left[\begin{array}{cc}
-g^{tt}+\left(1-\frac{1+\beta}{c_s^2}\right)(v_0^t)^2 & v_0^rv_0^t\left(1-\frac{1+\beta}{c_s^2}\right)\\
v_0^rv_0^t\left(1-\frac{1+\beta}{c_s^2}\right) & g^{rr}+\left(1-\frac{1+\beta}{c_s^2}\right)(v_0^r)^2
\end{array}\right]
\end{equation}
and 
\begin{equation}\label{G-lower-axi}
G_{\mu\nu}=-\left[\begin{array}{cc}
g^{rr}+\left(1-\frac{1+\beta}{c_s^2}\right)(v_0^r)^2 & -v_0^rv_0^t\left(1-\frac{1+\beta}{c_s^2}\right)\\
-v_0^rv_0^t\left(1-\frac{1+\beta}{c_s^2}\right) &  -g^{tt}+\left(1-\frac{1+\beta}{c_s^2}\right)(v_0^t)^2
\end{array}\right]
\end{equation}
where we neglected the conformal factors as earlier.

Eq. (\ref{G-upper-michel}), (\ref{metric_sph}), (\ref{G-upper-axi}) and (\ref{G-lower-axi}) our main results in this section and will be used for the study of acoustic spacetime in the following sections.
\section{Location of the acoustic horizon}
From Eq. (\ref{metric_sph}) and (\ref{G-lower-axi}) it is evident that the metric elements of the acoustic metric are time independent and hence the acoustic spacetime is stationary. In analogy to the general relativity, therefore, we can define the acoustic horizon as a time-like hypersurface whose normal $n_\mu$ is null with respect to the acoustic metric \cite{Poisson2004relativist,abraham:causal}
\begin{equation}
G^{\mu\nu} n_\mu n_\nu = 0
\end{equation}
The normal to any $r=\rm{constant}$ hypersurface is given by $\partial_\mu r=\delta^r_\mu$. Therefore the acoustic horizon is given by
\begin{equation}
G^{\mu\nu}\delta^r_\mu \delta^r_\nu = G^{rr}=0
\end{equation}
In analogy to the general relativity this acoustic horizon given by the above equation would represent  boundary of a region from which no phonons can escape to future null infinity. Causal structure of the acoustic spacetime, studied in Sec. \ref{sec:causal-structure}, illustrates this feature of the acoustic horizon.
\subsection{Michel Flow}
For Michel flow the acoustic horizon, given by $G^{rr}=0$, satisfies
\begin{equation}
-c_s^2+g_{rr}(v_0^r)^2(1-c_s^2)=0
\end{equation}
or \begin{equation}\label{horizon-michel-v}
c_s^2=\frac{g_{rr}(v_0^r)^2}{1+g_{rr}(v_0^r)^2}
\end{equation}
Now it is convenient to write Eq. (\ref{horizon-michel-v}) in terms of the advective  velocity $u_0$ defined in Eq. (\ref{u-in-co-rot-frame}). Eq. (\ref{v-t-in-co-rot-frame}) and (\ref{v-r-in-co-rot-frame}) give the expressions of $v^t$ and $v^r$ in terms of $u$ for spherically symmetric flow when $\lambda$ is set to zero. Thus
\begin{equation}\label{transform_sph}
v^t_0=\frac{1}{\sqrt{g_{tt}}}\frac{1}{\sqrt{1-u_0^2}}, \quad \quad v_0^r=\frac{u_0}{\sqrt{g_{rr}}}\frac{1}{\sqrt{1-u_0^2}} .
\end{equation}
Using these in Eq. (\ref{horizon-michel-v}) we find that the location of the acoustic event horizon is given by the relation 
\begin{equation}\label{sph-horizon}
{c_s}^2|_{\rm{h}}=u_0^2|_{\rm{h}}
\end{equation}
Hereafter the suffix `$\rm{h}$' will imply that the equation is to be evaluated at the horizon. The transonic surface  is also the surface where $|u_0|=c_s$. Thus for Michel flow the acoustic horizon and the transonic surface coincide.
\subsection{Axially symmetric flow}\label{horizon-loc-ax}
The acoustic event horizon satisfying $G^{rr}=0$ gives 
\begin{equation}
 c_s^2=\frac{(1+\beta)g_{rr}(v_0^r)^2}{1+g_{rr}(v_0^r)^2}
 \end{equation} 
Eq. (\ref{v-t-in-co-rot-frame}), (\ref{v-r-in-co-rot-frame}), (\ref{v-phi-in-co-rot-frame}) gives $v^t$, $v^r$ and $v^\phi$  in terms of $u$ and $\lambda$ as  \begin{equation}
 v^t_0=\sqrt{\frac{g_{\phi\phi}}{g_{tt}(g_{\phi\phi}-\lambda^2 g_{tt})}}\sqrt{\frac{1}{1-u_0^2}}
 \end{equation}
 \begin{equation}
 v_0^r=\frac{u_0}{\sqrt{g_{rr}(1-u_0^2)}}
 \end{equation}
 \begin{equation}
 v_0^\phi=\lambda \sqrt{\frac{g_{tt}}{g_{\phi\phi}(g_{\phi\phi}-\lambda^2 g_{tt})}}\sqrt{\frac{1}{1-u_0^2}}
 \end{equation}
Thus at the acoustic event horizon we have $c_s^2=(1+\beta)u_0^2$ or $u_0^2=\frac{c_s^2}{1+\beta}$. Where $\beta$ has the value as defined in Eq. (\ref{H1}). We thus observe that for axially symmetric accretion for flow in hydrostatic equilibrium along the vertical direction where $\beta\neq 0$, the effective sound speed becomes $c_s^{\rm eff} = \frac{c_s}{\sqrt{1+\beta}}$. $c_s^{\rm eff}$ is actually the speed of propagation of the linear perturbation inside the transonic fluid.

\section{Causal structure}\label{sec:causal-structure}
In this section we study the causal structure of the acoustic spacetime at and around the acoustic horizon to illustrate the behavior of the phonon null geodesics. The causal structure of the acoustic spacetime is independent of the conformal factor. The null geodesic corresponding to the radially traveling phonons is given by $ds^2=0$ with $\phi=$constant and $\theta=$constant. Thus we have 
\begin{equation}
(\frac{dr}{dt})_\pm\equiv b_\pm=\frac{-G_{rt}\pm \sqrt{G_{rt}^2-G_{rr}{G_{tt}}}}{G_{rr}}.
\end{equation}
So $t(r)$ is obtained as
\begin{equation}\label{b}
t(r)_\pm=t_0+\int_{r_0}^r \frac{1}{b_\pm}dr
\end{equation}
\subsection{Causal structure for Michel flow}
\begin{figure}[H]
\centering
\begin{tabular}{cc}
\resizebox{0.45\textwidth}{!}{\input{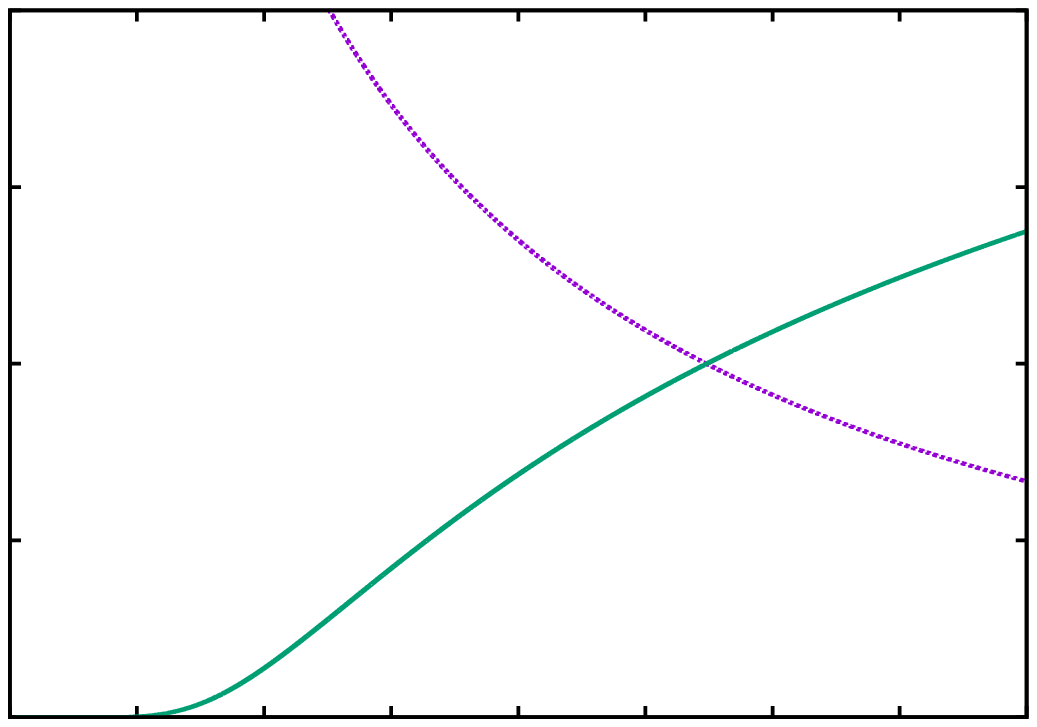}} &  \resizebox{0.45\textwidth}{!}{\input{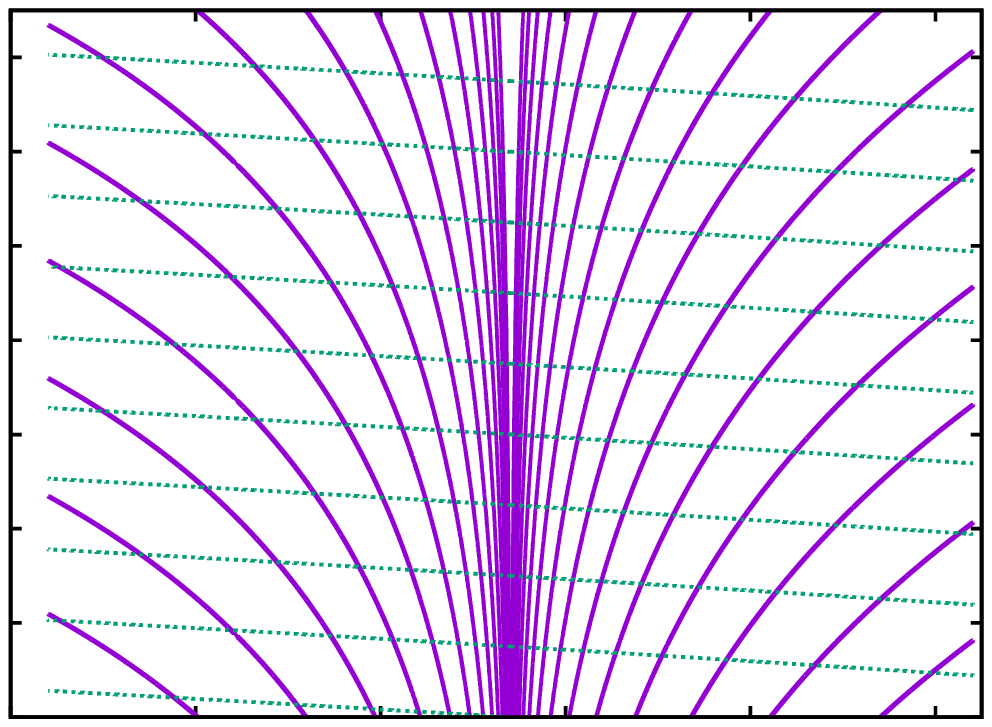}}\\
\end{tabular}
\caption{Left: Mach number vs radius plot of Michel flow. Right: Causal structure of the acoustic spacetime corresponding to the transonic flow (dashed line in the left figure). The dashed line is the  $t(r)_-$ vs $r$ and the solid line is $t(r)_+$ vs $r$. {where $t(r)_\pm$ is given by Eq. (\ref{b}).}}
\label{fig:1}
\end{figure}

For Michel flow the acoustic metric given by Eq. (\ref{metric_sph}). Using the expression of metric elements $G_{tt},G_{tr}$ and $G_{rr}$ and the Eq. (\ref{transform_sph}) gives
\begin{equation}
b_\pm=-\sqrt{\frac{g_{tt}}{g_{rr}}}\frac{(|u_0|\mp c_s)}{(1-|u_0|c_s)}
\end{equation}
where $g_{tt}$ and $g_{rr}$ are given by Eq. (\ref{metric_elements}). $b_-$ is always negative whereas $b_+$ is positive when $|u_0|<c_s$ i.e the flow is subsonic and it is negative when $|u_0|>c_s$ i.e when the flow becomes supersonic. We perform the integration given by Eq.  (\ref{b}) numerically using Euler method whereas the $b_\pm$ is obtained using 4$^{th}$ order Runge-Kutta method integrating the relativistic Euler equations for stationary flow. Below we plot $t(r)$ as a function of $r$ in Figure \ref{fig:1}.

From the causal structure of the acoustic spacetime shown in Figure \ref{fig:1} it is noticed that for $r>r_{\rm{h}}$, where for the current set up $r_{\rm{h}}\approx 274.08$, for increasing $t$, $r$ increases along the solid lines and decreases along the dashed line. Thus for $r>r_{\rm{h}}$ acoustic perturbations can move both outward and inward. However for $r<r_{\rm{h}}$, for increasing $t$, along both the lines, solid and dashed, $r$ decreases. Thus acoustic perturbations always move inward for $r<r_{\rm{h}}$. Therefore $r=r_{\rm{h}}$ represents an acoustic event horizon formed at $\mathcal{M}=1$ from which acoustic perturbations can not escape to the $r>r_{\rm{h}}$ region. Thus from the causal structure we can see that the transonic surface and the acoustic horizon coincides for isothermal Michel flow.

\subsection{Causal structure for axially symmetric flow}
\begin{figure}[] 
\centering
\begin{tabular}{cc}
\resizebox{0.45\textwidth}{!}{\input{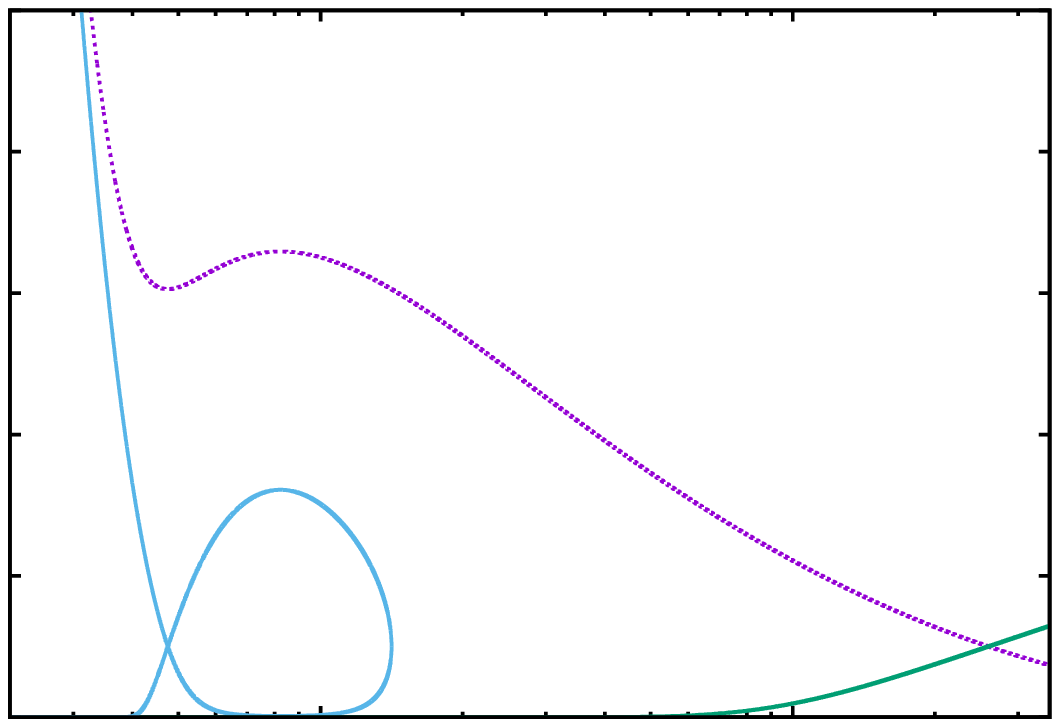}} & \resizebox{0.45\textwidth}{!}{\input{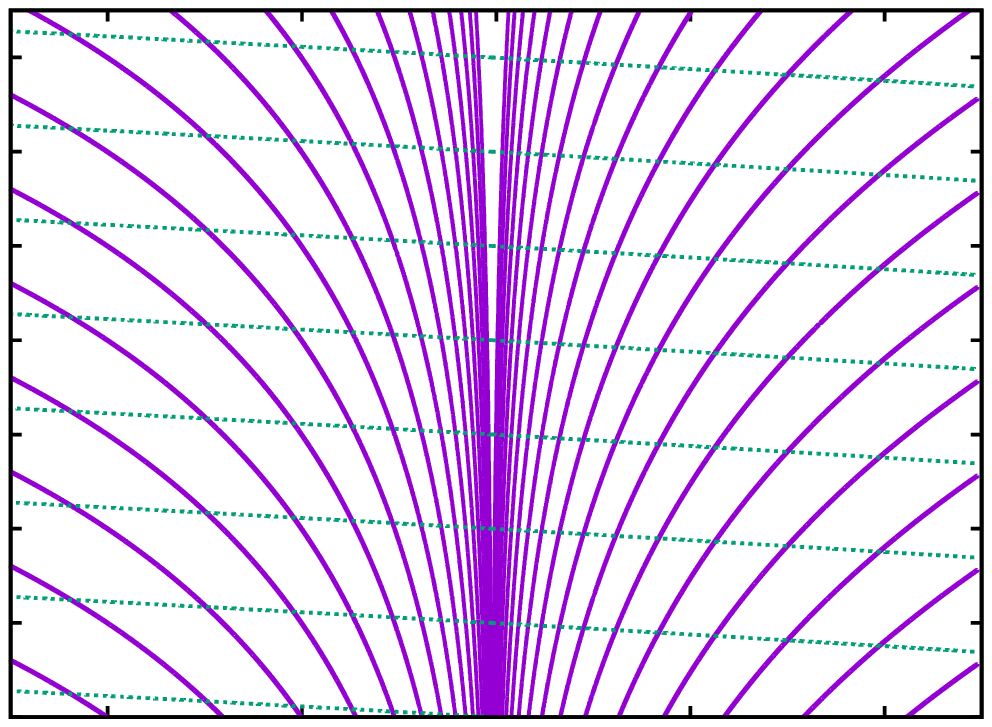}}\\
\resizebox{0.45\textwidth}{!}{\input{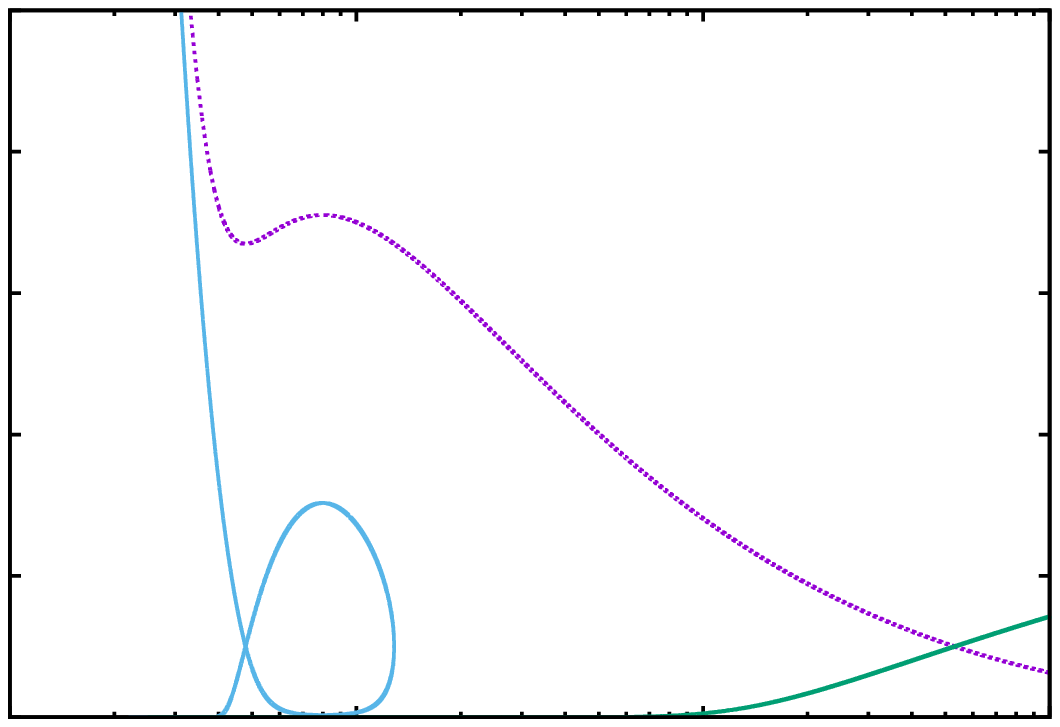}} & \resizebox{0.45\textwidth}{!}{\input{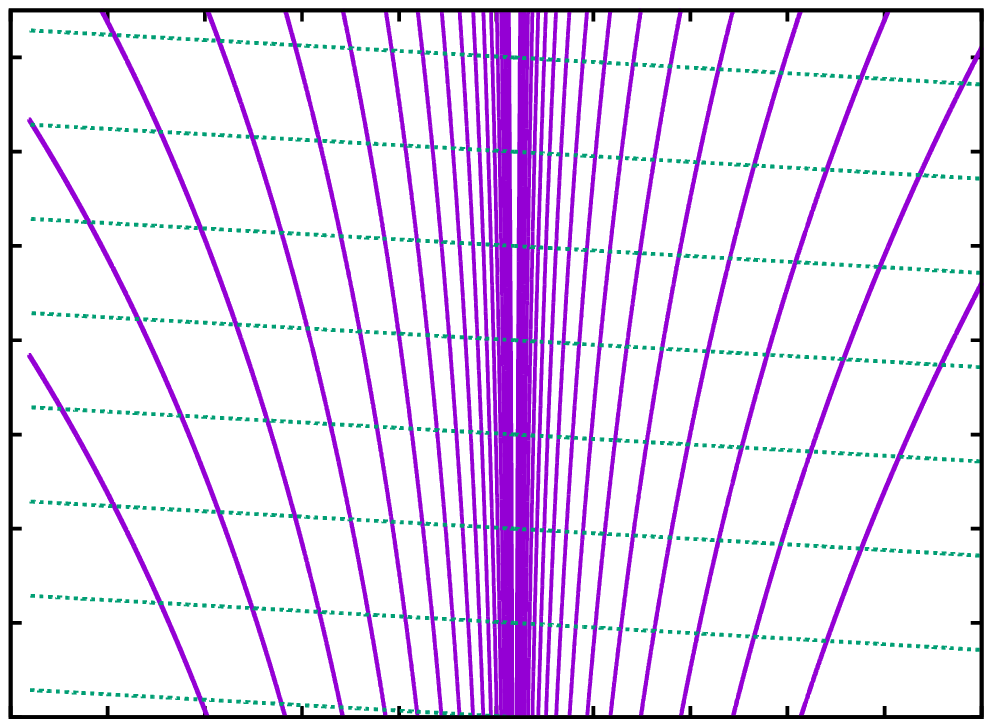}} \\
\resizebox{0.45\textwidth}{!}{\input{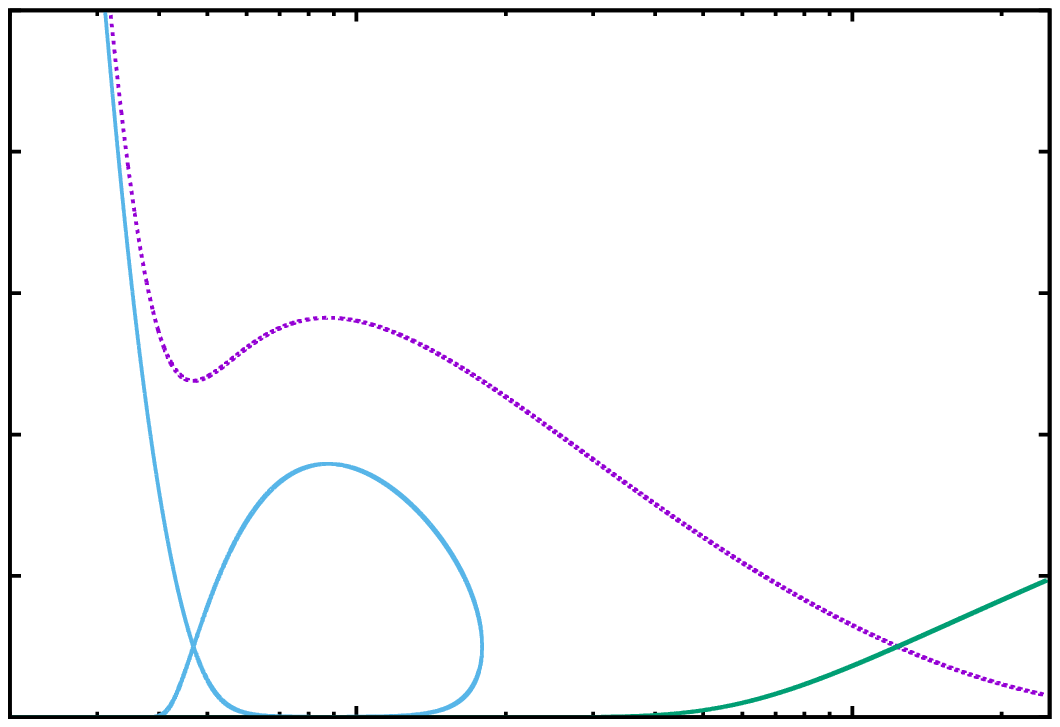}} & \resizebox{0.45\textwidth}{!}{\input{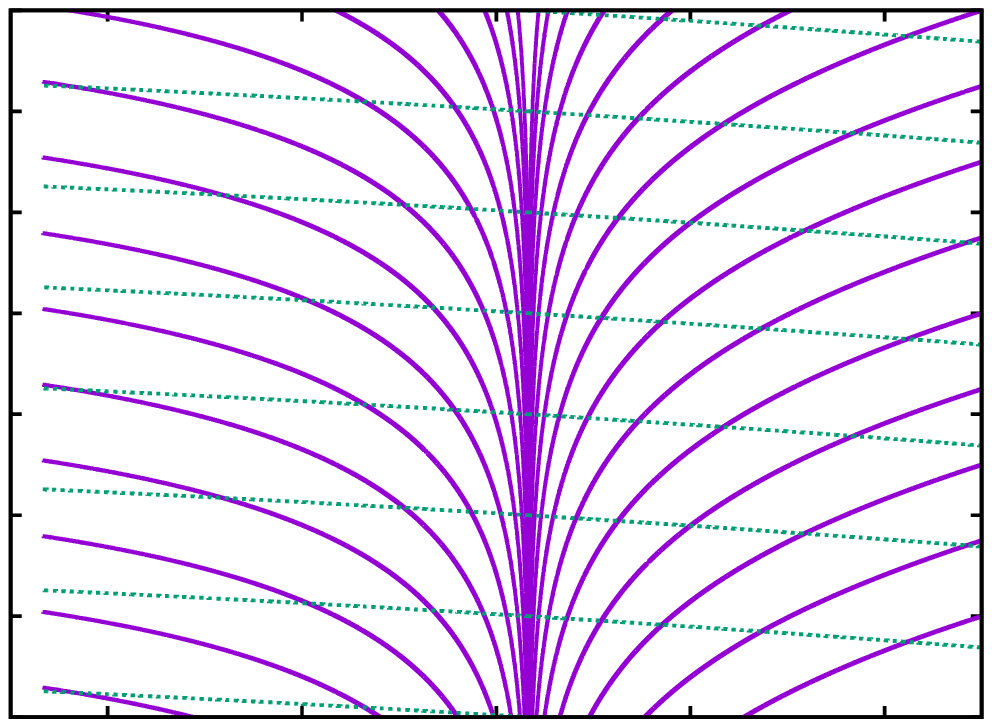}} 
\end{tabular}
\caption{We demonstrate the Mach number ($\mathcal{M}$) Vs. radial distance ($r$)
structure (phase portrait) for multi-transonic axisymmetric accretion and it's corresponding 
causal structures (at and around the sonic point(s)) for three different flow geometries, namely,
constant height flow (top panel), conical flow (middle panel) and flow in the 
hydrostatic equilibrium along the vertical direction (lower panel). The isothermal 
accretion has been characterized by $\lambda=3.75$ and $T=10^{10}K$. The panels show how the location of acoustic horizon and the curvature at and around the horizon varies for different flow geometries. On the right the dashed line is the  $t(r)_-$ vs $r$ and the solid line is $t(r)_+$ vs $r$. where $t(r)_\pm$ is given by Eq. (\ref{b}).}\label{fig:2}
\end{figure}

For axially symmetric flow the stationary flow configuration depends on the specific angular momentum $\lambda$. Thus the Mach number $\mathcal{M}$ vs $r$ plot as well as the causal structure will depend on the set of parameters $[T,\lambda]$. Different set of values of these parameters allows different kind of $\mathcal{M}$ vs $r$ plots. One can plot the parameter space $[T,\lambda]$ and identify the regions which allows single critical point or multiple critical points. Where critical points are the roots of the polynomial obtained by setting the denominator, in the analytic expression for $\frac{du_0}{dr}$, to zero. However this is beyond the scope of this work and we refer for details to \cite{Tarafdar-das-2015,abraham:causal}. To demonstrate the characteristic features of the $\mathcal{M}$ vs $r$ plot and the causal structure we have chosen a particular values of the parameters $[T,\lambda]$ which allows multiple critical points and such that the accretion flow passes through the outer critical point. For axially symmetric flow $G_{tt},G_{tr},G_{rr}$ are given by
\begin{eqnarray}
G_{tt}=\frac{(1+\beta)u_0^2-c_s^2}{c_s^2(1-u_0^2){(1-2/r)^{-1}}}\\
G_{tr}=G_{rt}=\frac{|u_0|(1+\beta-c_s^2)\sqrt{\frac{r^2}{r^2-\lambda^2(1-2/r)}}}{c_s^2(1-u_0^2)}\\
G_{rr}=\frac{c_s^2(1-u_0^2)+(1+\beta-c_s^2)\frac{r^2}{r^2-\lambda^2(1-2/r)}}{c_s^2(1-u_0^2)(1-2/r)}
\end{eqnarray}
with $\beta=0$ for conical and constant height flow and $\beta=c_s^2$ for flow with hydrostatic equilibrium in vertical direction. Similar to Michel flow we perform the integration given by (\ref{b}) numerically using Euler method whereas the $b_\pm$ is obtained using 4$^{th}$ order Runge-Kutta method integrating the relativistic Euler equations for stationary flow. We plot $t(r)$ as a function of $r$ for different flow geometries in Figure \ref{fig:2}.

In the present work we have studied the stationary transonic integral solutions of accretion flow containing more than one critical points to understand the causal structure of the embedded relativistic acoustic geometry. We, however, did not probe the shock formation phenomena in this work. If shock forms, an analogue white hole is expected to appear at the shock location and it should show up in the causal structure \cite{abraham:causal}.

\section{Surface Gravity}
The acoustic metric $G_{\mu\nu}$ are identical for each of the three perturbation up to a conformal factor which does not affect our results \cite{conformal_invariance}. Therefore we get the same acoustic surface gravity for the acoustic metric obtained by perturbing three different quantities for a particular flow geometry as discussed below.
\subsection{Surface gravity for Michel flow}
The acoustic metric $G_{\mu\nu}$ is independent of time. Therefore we have the stationary Killing vector $\chi^\mu=\delta^\mu_t$. Further we see that the Killing vector is null at horizon and thus Killing horizon coincides with the event horizon. The norm of Killing vector is $\chi^\mu \chi_\mu=G_{tt}$ which is zero at horizon as
\begin{equation}
\frac{-c_s^2+\frac{u^2}{1-u^2}(1-c_s^2)}{g_{rr}}|_{u_0^2=c_s^2}=0
\end{equation}
Therefore the surface gravity can be evaluated using the Killing vector $\chi^\mu$ using the following definition \cite{Wald,Poisson2004relativist,Bilic1999}
\begin{equation}
\nabla_\alpha \left( -\chi^\mu \chi_\mu\right)=2\kappa \chi_\alpha
\end{equation}
where $\kappa$ is the surface gravity. It is important to keep in mind that the above equation is to be evaluated on the horizon. Taking $\alpha=r$ we can get $\kappa$ from the above equation. $\chi_\alpha=G_{\alpha \mu}\chi^\mu=G_{\alpha t}$, so $\chi_r=G_{rt}$. So $\kappa$ is given by
\begin{equation}
\kappa=\frac{1}{2}\frac{1}{G_{rt}}\partial_r(-G_{tt})|_{u_0^2=c_s^2}
\end{equation}
So the acoustic surface gravity in terms of background metric elements and velocity is given by
\begin{equation}
\kappa=\left|\frac{\sqrt{\frac{g_{tt}}{g_{rr}}}}{1-c_s^2}\frac{du_0}{dr}\right|_{\rm{h}}
\end{equation}

\subsection{Surface gravity for axially symmetric flow}
Similar to the spherically symmetric here also we have the stationary Killing vector $\chi^\mu=\delta^\mu_t$ which is null at the horizon where $c_s^2=(1+\beta)u^2$. So the surface gravity is given by the formula 
\begin{equation}
\kappa=\frac{1}{2}\frac{1}{G_{rt}}\partial_r (-G_{tt})|_{(1+\beta)u_0^2=c_s^2}
\end{equation}
After a simple calculation the acoustic surface gravity for axially symmetric flow in terms of background metric and velocity is given by
\begin{equation}
\kappa=\left|\frac{1+\beta}{1+\beta-c_s^2}\sqrt{\frac{g_{tt}}{g_{rr}g_{\phi\phi}}(g_{\phi\phi}-\lambda^2 g_{tt})}\frac{du_0}{dr}\right|_{\rm{h}}
\end{equation}
The expression of the acoustic surface gravity for schwarzschild background becomes
\begin{equation}
\kappa=\left|\frac{(1+\beta)}{(1+\beta-c_s^2)}\frac{(r-2)}{r^2}\sqrt{r^2-\lambda^2(1-\frac{2}{r})}\frac{du_0}{dr}\right|_{\rm{h}}
\end{equation}
where $\beta$ has different values for different flow geometry as mentioned in \ref{Liner-perturbation-axial}. Hence the above result shows how the acoustic surface gravity for axially symmetric flow depends on the specific angular momentum $\lambda$ and the flow geometry. Further the equation is to be evaluated at the acoustic horizon which as indicated in Sec. \ref{sec:causal-structure} depends sensitively on the flow geometry.
\section{Concluding Remarks}\label{Concluding-Remarks}
One of the major findings of our work presented in this paper is to be realized as follows: The salient features of the embedded acoustic geometry do not depend on the physical quantity we linearly perturb to obtain the sonic metric. Linear perturbation of the velocity potential, the relativistic Bernoull's constant and the mass accretion rate provide the same acoustic metric, apart from a conformal factor. This conformal factor does not contribute to the acoustic horizon as well as the value of the acoustic surface gravity. The acoustic geometry is thus an intrinsic property of the accreting black hole system. At the same time, what we find is that for the same set of initial boundary conditions used  (to describe the fluid system), the location of the acoustic horizon as well as the value of the surface gravity depends, quite sensitively, on the geometric configuration of matter - different disc models provide different values of $r_{\rm{h}}$ and $\kappa$ for same set of values $[T,\lambda]$. Hence the properties of the embedded acoustic metric does not depend on the physical quantity we perturb to obtain the metric, rather such properties get influenced by the matter geometry in connection to the axially symmetric accretion.

In Sec. \ref{sec:causal-structure}, it has explicitly been demonstrated how the location of the acoustic metric may be determined. For transonic accretion, one can find the location of the transonic point through critical point analysis of the space derivatives of the integral stationary solutions. The accreting matter makes a transition from the subsonic to supersonic state at such transonic point- such transitions are smooth and continuous. For shock, the sonicity changes, discontinuously, the reverse order-here the supersonic flow becomes subsonic. It has been argued that the transonic points are the location of the acoustic black holes, whereas an acoustic white hole forms at the shock location \cite{abraham:causal}. In this work we have formally established such identification of the acoustic horizons with the transonic points through the construction of the causal structure at and around the sonic points for stationary transonic integral flow solutions. The phonons, i.e., the null geodesic for this problem are studied through such constructions for axially symmetric flows in three different geometric configuration of matter.

Since, in connection to the analogue space time, the
sound wave is defined as the low amplitude linear perturbation \cite{Visser1998}, we concentrate on the linear perturbation scheme only. It will, however, be tempting to investigate whether the emergence of the analogue space time is associated only with the linear perturbation of the background flow, or in general, even for non linear perturbation, one observes the formation of the acoustic space time. It has recently been found that for non linear or higher order perturbation within Newtonian frame work, or perturbation with large amplitude, one may not obtain stationary static horizons \cite{arnab-nonlin-sph,arnab-nonlin-ax,mach-nonlin}. In our future works, we would like to study the consequence of non linear higher order perturbations of relativistic accretion. Such work is in progress and will be reported elsewhere.

The stationary integral solution of the relativistic Euler equation provides the relativistic {Bernoulli's}
constant. The velocity potential, however, is introduced through the irrotationality criteria. The irrotationality condition is required to derive the relativistic Euler equation, from which one obtains the {Bernoulli's}
constant as one of the two first integrals of motion of the system. Hence the conformal factor corresponding
to the acoustic metric obtained by perturbing the velocity potential and the {Bernoulli's} constant are exactly the same, where as it is different for the acoustic metric obtained through the perturbation of the mass accretion rate, since the mass accretion rate is the stationary integral solution of the continuity equation and the irrotationality condition does not play any role in formulating the mass conservation equation.

In this work, we have computed the value of the acoustic surface gravity for spherical as well as axial
symmetric black-ground flow structure for various geometric configurations of matter. One needs to know the
values of the unperturbed values of the bulk dynamical velocity, the speed of propagation of sound, and the
space derivative of the bulk velocity - all evaluated at the acoustic horizon, to calculate the exact numerical
value of the acoustic surface gravity. The parameters governing the flow, i.e., the conserved specific angular
momentum as well as the flow temperature is also necessary to know. This is possible when $[u, c_s , du/dr]$ can be computed for stationary integral flow solutions at the sonic point. Such values can be obtained by numerically integrating the flow solutions for the corresponding phase orbits \cite{Tarafdar2016}. For axially symmetric flows, there might be multiple sonic points and hence the same stationary background flow can provide multiple values of the acoustic surface gravity. Also, stationary shock may form at the axially symmetric
flow (black hole accretion discs), where instead of a black hole, an acoustic white hole can be generated, and
one can show that the acoustic surface gravity can assume infinitely large values at the shock for inviscid
flows. For viscous transonic accretion, however, one can have a very large yet finite value of the acoustic
surface gravity, although viscosity may destroy the Lorenzian symmetry and a modified approach for the
perturbation technique is required to address this issue - such discussion is beyond the scope of the present
work and will be reported elsewhere.

The usual role of the surface gravity is to serve as the gateway to calculate the acoustic Hawking like
temperature. However, recent works shows that that is not the only way to extract further information
about the system from the value of the surface gravity, rather the acoustic surface gravity itself is a rather
significant entity to realize features of the sonic metric, irrespective of the existence of Hawking like phenomena based on the quantum field theoretic formalism in curved space time. For an analogue system, through
the anomalous dispersion relation, {$\kappa$} plays an important role in understanding the non-universality of the
Hawking like effects, where the deviation of the Hawking like effects depends (at very sensitive scale) on the
space gradient of the bulk velocity of the background fluid flow (see, e.g., the comprehensive review article
\cite{Robertson2012} and references therein). In our calculation, $\kappa$ contains $du/dr$ and hence we provide a solution scheme where the aforementioned non universality may actually be numerically quantified, through observationally estimated physical variables, for a large scale gravitating system.

We have studied the accretion flow in the Schwarzschild metric. The astronomers, however, believe that the majority of the astrophysical black holes posses spin angular momentum \cite{Miller2009,Kato2010,Ziolkowski2010,Tchekhovskoy2010,
Daly2011,Buliga2011,Reynolds2012,McClintock2011,Martínez-Sansigre2011,Dauser2010,Nixon2011,McKinney2012,
McKinney2013,Brenneman,Rolando2013,Sesana,Fabian2014,
Healy,Jiang,Nemmen}. This tempted us to perform our calculation for astrophysical accretion in the Kerr metric to observe the variation of the
properties of the sonic geometry with the black hole spin parameter. Such work is in progress and will be
reported shortly.

\ack
Visit of IF has been supported by the XII$^{\rm{th}}$ plan budget of Cosmology and High Energy Astrophysics grant. Authors acknowledge useful discussions with S. Datta, P. Tarafdar, D. A. Bollimpalli and S. Nag.

\appendix 
\section{Irrotationality condition for isothermal flow in curved background} \label{Appendix:irrotationality-condition}
Vorticity in general relativity is defined as
\begin{equation}
\omega_{\mu\nu}=l^\rho_{~\mu} l^\sigma_{~\nu} v_{[\rho;\sigma]}
\end{equation}
where \begin{equation}
l^\mu_{~\nu}=\delta^\mu_{~\nu}+v^\mu v_\nu \quad \rm{and}\quad v_{[\mu;\nu]}=\frac{1}{2}[\nabla_\nu v_\mu-\nabla_\mu v_\nu]
\end{equation}
Thus $\omega_{\mu\nu}$ after some calculation can be written as
\begin{equation}\label{omega1}
 \omega_{\mu\nu}=v_{[\mu;\nu]}+\frac{1}{2}v^\rho[v_\nu\nabla_\rho v_\mu-v_\mu \nabla_\rho v_\nu]
 \end{equation} 
 The relativistic Euler Eq. (\ref{Eq:euler-isothermal}) gives
 \begin{equation}
 v^\mu \nabla_\mu v_\nu=-\frac{c_s^2}{\rho}[\partial_\nu \rho+v^\mu v_\nu \partial_\mu \rho]
 \end{equation}
 Using above equation in Eq. (\ref{omega1}) and rearranging finally gives 
 \begin{equation}
 \omega_{\mu\nu}=\frac{1}{2\rho^{c_s^2}}[\partial_\nu(v_\mu \rho^{c_s^2})-\partial_\mu(v_\nu \rho^{c_s^2})]
 \end{equation}
 Thus irrotationality condition becomes
 \begin{equation}
 \partial_\mu(v_\nu \rho^{c_s^2})-\partial_\nu(v_\mu \rho^{c_s^2})=0
 \end{equation}

\section{Linear perturbation analysis and acoustic metric for Michel flow}\label{Linear-perturbation-michel}
In this section we demonstrate how to obtain the wave equation of propagation of perturbations when we linearly perturb the velocity potential, relativistic Bernoulli's constant and the mass accretion rate for a spherically symmetric Michel flow. For such flow the velocity components $v^\theta$ and $v^\phi$ are zero and the normalization condition given by $v^\mu v_\mu=-1$ gives the relation between $v^t$ and $v^r$ as
\begin{equation}
v^t=\sqrt{\frac{1+g_{rr}(v^r)^2}{g_{tt}}}
\end{equation}
\subsection{Perturbation of velocity potential field}\label{Linear-pert-velo-michel}
In \cite{Bilic1999} acoustic metric for a general background metric $g_{\mu\nu}$ was derived by perturbing the potential field $\psi$, defined as $h v_\mu = -\partial_\mu \psi$, for adiabatic flow.  Where a flow is called adiabatic when the total heat content of the fluid system is conserved and is governed by the equation of state $p\propto \rho^\gamma$, where $\gamma$ is the specific heat of the fluid. For such flow the sound speed is given as $\sqrt{\frac{\partial p}{\partial \epsilon}}$. $h, p, \rho$ and $\epsilon$ have the usual meanings as defined in Sec. \ref{Sec:methodology}. 

For isothermal flow we introduced a potential field in the flow, in a similar manner,  using irroationality condition for isothermal flow given by Eq. (\ref{irrotationality}) in Sec. \ref{Sec:Definition-velocity-potential}. The potential field $\psi$ is defined in Eq. (\ref{definition-velocity-potential}).  We now derive the acoustic metric for a general background metric $g_{\mu\nu}$ by introducing 
\begin{eqnarray}\label{velocity perturbation }
\rho=\rho_0+\rho_1\\ \nonumber
v^\mu=v^\mu_0+v^\mu_1\\ \nonumber
\psi=\psi_0+\psi_1 \nonumber
\end{eqnarray}
where $\rho_1,v^\mu_1$ and $\psi_1$ are small acoustic disturbances around some average bulk motion represented by $\rho_0,v^\mu_0$ and $\psi_0$. The normalization condition $v^\mu v_\mu=-1$ gives 
\begin{equation}
g_{\mu\nu}v^\mu v^\nu_1=0
\end{equation}
substituting $\rho,v^\mu,\psi$ using Eq. (\ref{velocity perturbation }) in Eq. (\ref{definition-velocity-potential}) and retaining only terms which are linear in perturbations gives 
\begin{equation}\label{velocity rho1}
\rho^{c_s^2-1}c_s^2\rho_1=v^\mu_0 \partial_\mu \psi_1
\end{equation}
and 
\begin{equation}\label{velocity v1}
\rho_0^{c_s^2}v^\mu_1=-g^{\mu\nu}\partial_\nu \psi_1-v^\mu_0 v^\nu_0\partial_\nu \psi_1
\end{equation}
Substituting $\rho,v^\mu,\psi$ using Eq. (\ref{velocity perturbation }) in the  continuity equation given by Eq. (\ref{cont eq}) and keeping terms that are linear in perturbations gives
\begin{equation}\label{velo cont pert}
\partial_\mu \left[\sqrt{-g}\left(v^\mu \rho_1+\rho_0 v^\mu_1\right)\right]=0
\end{equation}
Now using $\rho_1$ and $v^\mu_1$ form Eq. (\ref{velocity rho1}) and (\ref{velocity v1}) respectively in the above equation gives 

\begin{equation}
\partial_\mu [\frac{\sqrt{-g}}{\rho^{c_s^2-1}}\{g^{\mu\nu}+(1-\frac{1}{c_s^2})v^\mu_0v^\nu_0\}\partial_\nu \psi_1]=0
\end{equation}
The above equation can be written as $\partial_\mu(f^{\mu\nu}\partial_\nu \psi_1)=0$ where $f^{\mu\nu}$ is 
\begin{equation}
f^{\mu\nu}=k_1(r)\left[-c_s^2g^{\mu\nu}+\left(1-c_s^2\right)v^\mu_0v^\nu_0\right]
\end{equation}
where $k_1(r)=-(\sqrt{-g})/(\rho^{c_s^2-1}c_s^2)$. 

In the above derivation we have not considered any symmetry of the flow or the background metric. For spherically symmetric radial accretion in Schwarzschild background Eq. (\ref{velocity rho1}) gives

\begin{equation}\label{velo rho1 sph}
\rho_1=\frac{1}{\rho_0^{c_s^2-1}c_s^2}(v_0^t \partial_t \psi_1+v_0^r\partial_r\psi_1)
\end{equation}
and Eq. (\ref{velocity v1}) gives 
\begin{eqnarray}\label{velo vt1 sph}
v_1^t=\frac{1}{\rho_0^{c_s^2}}[(g^{tt}-(v_0^t)^2)\partial_t\psi_1-v_0^rv_0^t\partial_r\psi_1]\\
v_1^r=\frac{1}{\rho_0^{c_s^2}}[(-g^{rr}-(v_0^r)^2)\partial_r\psi_1-v_0^rv_0^t\partial_t\psi_1] \label{velo v1 sph}
\end{eqnarray}
Eq. (\ref{velo cont pert}) becomes
\begin{equation}
\partial_t[\sqrt{-g}(v_0^t\rho_1+\rho_0v_1^t)]+\partial_r[\sqrt{-g}(v_0^r\rho_1+\rho_0v_1^r)]=0
\end{equation}
Substituting $\rho_1,v^t_1$ and $v_1^r$ in the above equation using equations (\ref{velo rho1 sph}) (\ref{velo vt1 sph}) and (\ref{velo v1 sph}) respectively gives  

\begin{eqnarray}
\fl
\partial_t[k_1(r)(c_s^2g^{tt}+(1-c_s^2)(v_0^t)^2)\partial_t\psi_1]+\partial_t[k_1(r)(1-c_s^2)v_0^rv_0^t \partial_r\psi_1] \\ \nonumber
+\partial_r[k_1(r)(1-c_s^2)v_0^rv_0^t \partial_t\psi_1]+\partial_r[k_1(r)(-c_s^2g^{rr}+(1-c_s^2)(v_0^r)^2)\partial_r\psi_1]=0
\end{eqnarray}
Thus the $f^{\mu\nu}$ becomes
\begin{equation}
f^{\mu\nu}=k_1(r)\left[\begin{array}{cc}
c_s^2g^{tt}+(1-c_s^2)(v_0^t)^2 & (1-c_s^2)v_0^rv_0^t \\
(1-c_s^2)v_0^rv_0^t & -c_s^2g^{rr}+(1-c_s^2)(v_0^r)^2

\end{array}\right]
\end{equation}
\subsection{Perturbation of the relativistic Bernoulli's constant}\label{Linear-pert-bern-michel}
In this section we do the linear perturbation analysis for relativistic Bernoull's constant. In Eq.  \ref{Define-xi} we defined the quantity $\xi(r,t)=v_t(r,t)\rho^{c_s^2}(r,t)$  which has the stationary value equal to $\xi_0=\rm{const.}$ We perturb $\rho(r,t),v^r(r,t)$ and $v^t(r,t)$ up to linear order about their background stationary values as following
\begin{eqnarray}\label{bernoulli pert spherical}
\rho(r,t)=\rho_0(r)+\rho_1(r,t)\\ \nonumber
v^r(r,t)=v_0^r(r)+v_1^r(r,t)\\  \nonumber
v^t(r,t)=v^t_0(r)+v^t_1(r,t) \nonumber
\end{eqnarray}
where $\rho_1(r,t),v_1^r(r,t),v^t_1(r,t)$ are the first order perturbed terms. Eq. \ref{Define-xi} gives $\xi(r,t)=\xi_0+\xi_1(r,t)$. In terms of $\xi$ Eq. (\ref{time comp of ber eq sph}) can be recast in the form 
\begin{equation}\label{time comp ber in xi sph}
v^t\partial_t v^t+\frac{c_s^2}{\rho}\frac{g_{rr}(v^r)^2}{g_{tt}}\partial_t\rho+\frac{v^rv^t}{\xi}\partial_r\xi=0
\end{equation}
using the equations (\ref{bernoulli pert spherical}) in Eq. (\ref{time comp ber in xi sph}) and collecting only the terms which are linear in the perturbation terms gives 

\begin{equation}\label{drxi sph}
-\frac{\alpha \xi_0}{v_0^r}\partial_t v_1^r-\frac{c_s^2}{\rho_0}\alpha \xi_0 \partial_t \rho_1=\partial_r \xi_1
\end{equation}
where we have used  
\begin{equation}
v^t_1=\alpha(r) v_1^r
\end{equation}
with $\alpha(r)=(g_{rr}v_0^r)/(g_{tt}v_0^t)$. Here the symbol $\alpha$ has been used to denote the collection of few algebraic terms. $\alpha$ has no physical significance as such, it has just been used to conveniently express the aforementioned collection of terms.  We also have 
\begin{equation}\label{xi1 ber sph}
\xi_1=\frac{\xi_0\alpha}{v_0^t} v_1^r+\frac{\xi_0 c_s^2}{\rho_0} \rho_1.
\end{equation}
Differentiating both sides with respect to $t$ gives 
\begin{equation}\label{dtxi sph}
\frac{\xi_0\alpha}{v_0^t}\partial_t v_1^r+\frac{\xi_0 c_s^2}{\rho_0}\partial_t \rho_1=\partial_t \xi_1
\end{equation}
using Eq. (\ref{drxi sph}) and (\ref{dtxi sph}) $\partial_t v_1^r$ and $\partial_t \rho_1$ can be written solely in terms of $\partial_t \xi_1$ and $\partial_r \xi_1$
\begin{equation}\label{dtv1 ber sph}
\partial_t v_1^r=\frac{1}{\Delta}\frac{\xi_0 c_s^2}{\rho_0}\left[\partial_r \xi_1+\alpha\partial_t\xi_1 \right]
\end{equation}
and \begin{equation}\label{dtrho1 ber sph}
\partial_t\rho_1=-\frac{1}{\Delta}\frac{\xi_0 \alpha}{v_0^rv_0^t}\left[ v_0^r\partial_r\xi_1+v_0^t \partial_t\xi_1 \right]
\end{equation}
where $\Delta=(-\xi_0^2 c_s^2\alpha)/(\rho_0 v_0^rg_{tt}(v_0^t)^2)$. Using equations (\ref{bernoulli pert spherical}) in the continuity equation given by Eq. (\ref{cont eq in sph}) and retaining only the terms which are of first order in the perturbation terms, provides
\begin{equation}
\partial_t \left[ \sqrt{-g}(\rho_1 v_0^t+\alpha \rho_0 v_1^r)\right]+\partial_r \left[ \sqrt{-g}(\rho_1 v^r_0+\alpha \rho_0 v_1^r)\right]=0
\end{equation}
differentiating both sides of the above equation and substituting $\partial_t v_1^r$ and $\partial_t \rho_1$ using Eq. (\ref{dtv1 ber sph}) and (\ref{dtrho1 ber sph}) respectively gives
\begin{eqnarray}
\fl
\partial_t[k_2(r)\frac{c_s^2+g_{tt}(v_0^t)^2(1-c_s^2)}{g_{tt}} \partial_t\xi_1]+\partial_t\left[k_2(r)v^r_0v_0^t(1-c_s^2)\partial_r \xi_1\right]\\ \nonumber
+\partial_r\left[k_2(r)v^r_0v_0^t(1-c_s^2)\partial_t \xi_1\right]\\ \nonumber
+\partial_r[k_2(r)\frac{-c_s^2+g_{rr}(v^r_0)^2(1-c_s^2)}{g_{rr}} \partial_r\xi_1]=0
\end{eqnarray}
The above equation is of the form $\partial_\mu \left(f^{\mu\nu}\partial_\nu \xi_1\right)=0$. So we get $f^{\mu\nu}$ to be 
\begin{equation}
f^{\mu\nu}=k_2(r)\left[\begin{array}{ccc}
\frac{c_s^2+g_{tt}(v_0^t)^2(1-c_s^2)}{g_{tt}} & v^r_0v_0^t(1-c_s^2)\\
v^r_0v_0^t(1-c_s^2) & \frac{-c_s^2+g_{rr}(v^r_0)^2(1-c_s^2)}{g_{rr}}
\end{array}\right]
\end{equation}
where $k_2(r)=-(\sqrt{-g})/(c_s^2\rho^{c_s^2-1})$.
\subsection{Perturbation of the Mass Accretion Rate}\label{Linear-pert-mass_michel}
In this section we obtain the acoustic metric  by perturbing the mass accretion rate. Using $\xi_1$ given by Eq. (\ref{xi1 ber sph}) in the Eq. (\ref{drxi sph}), obtained by using the perturbation quantities in the time component of the Euler equation, we have
\begin{eqnarray}\label{pert mass sph}
\frac{\alpha}{v_0^r}\partial_tv_1^r+\frac{c_s^2}{\rho_0}\alpha \partial_t\rho_1+\partial_r(\frac{\alpha}{v_0^t}v_1^r+\frac{c_s^2}{\rho_0}\rho_1 )=0
\end{eqnarray}
The Eq. (\ref{Define-mass}) defines a variable $\Psi=\sqrt{-\tilde{g}}v^r(r,t)\rho(r,t)$ which has the stationary value equal to the mass accretion rate $\Psi_0$ such that $\Psi(r,t)=\Psi_0+\Psi_1(r,t)$. Using Eq. (\ref{bernoulli pert spherical}) we get 
\begin{equation}\label{psi1 mass sph }
\Psi_1(r,t)=\sqrt{-\tilde{g}}(v^r_0\rho_1+\rho_0 v_1^r)
\end{equation}
Again using Eq. (\ref{bernoulli pert spherical}) in the continuity equation given by (\ref{cont eq in sph}) we get
\begin{equation}\label{drpsi1 mass sph}
\partial_t[v_0^t \rho_1+\frac{g_{rr}v^r_0}{g_{tt}v_0^t}\rho_0 v_1^r]+\frac{1}{\sqrt{-\tilde{g}}}\partial_r\Psi_1=0
\end{equation}
using Eq. (\ref{psi1 mass sph }) and (\ref{drpsi1 mass sph}) we can express $\partial_t v_1^r$ and $\partial_t\rho_1$ solely in terms of $\partial_t \Psi_1$ and $\partial_r\Psi_1$, 
\begin{equation}\label{dtv1 mass sph}
\partial_t v_1^r=\frac{g_{tt}v_0^t}{\rho_0\sqrt{-\tilde{g}}}\left[ v_0^t \partial_t \Psi_1+v^r_0\partial_r\Psi_1\right]
\end{equation}
and 
\begin{equation}\label{dtrho1 mass sph}
\partial_t\rho_1=-\frac{1}{\sqrt{-\tilde{g}}}\left[g_{rr}v^r_0\partial_t\Psi_1+g_{tt}v_0^t\partial_r\Psi_1\right]
\end{equation}
Now taking a time derivative of the Eq. (\ref{pert mass sph}) and substituting $\partial_tv_1^r$ and $\partial_t\rho_1$ using (\ref{dtv1 mass sph}) and (\ref{dtrho1 mass sph}) we obtain
\begin{eqnarray} \label{mass_michel_f_eq}
\fl
\partial_t[k_3(r)\frac{c_s^2+g_{tt}(v_0^t)^2(1-c_s^2)}{g_{tt}} \partial_t\Psi_1]+\partial_t\left[k_3(r)v^r_0v^t_0(1-c_s^2)\partial_r \Psi_1\right]\\ \nonumber
+\partial_r\left[k_3(r)v^r_0v^t_0(1-c_s^2)\partial_t \Psi_1\right]\\ \nonumber
+\partial_r[k_3(r)\frac{-c_s^2+g_{rr}(v^r_0)^2(1-c_s^2)}{g_{rr}} \partial_r\Psi_1]=0
\end{eqnarray}
As before this is of the form $\partial_\mu\left( f^{\mu\nu}\partial_\nu \Psi_1\right)=0$. So we get $f^{\mu\nu}$ to be
\begin{equation}
f^{\mu\nu}=k_3(r)\left[\begin{array}{cc}
\frac{c_s^2+g_{tt}(v_0^t)^2(1-c_s^2)}{g_{tt}} & v^r_0v^t_0(1-c_s^2)\\
v^r_0v^t_0(1-c_s^2) & \frac{-c_s^2+g_{rr}(v^r_0)^2(1-c_s^2)}{g_{rr}}
\end{array}\right]
\end{equation}
where $k_3(r)=-(g_{rr}v^r_0)/(v_0^t)$.

\section{Linear perturbation analysis and acoustic metric for Axially Symmetric Flow}\label{Liner-perturbation-axial}
We now consider an axially symmetric disc like flow. There are three geometries that we can consider. These are disc with constant thickness, conical flow and disc under hydrostatic equilibrium in the vertical direction as mentioned earlier. These models are discussed in Sec. \ref{Accreting black holes as analogue systems}. We shall work with these three model in a unified way. The normalization condition $v^\mu v_\mu=-1$ gives 
\begin{equation}\label{normalization axi}
g_{tt}(v^t)^2=1+g_{rr}(v^r)^2+g_{\phi\phi}(v^\phi)^2
\end{equation}
From irrotationality condition given by Eq. (\ref{irrotationality}) with $\mu=t$ and $\nu=\phi$ and with axially symmetry we have 
\begin{equation}
\partial_t(v_\phi\rho^{c_s^2})=0
\end{equation}
again with $\mu=r$ and $\nu=\phi$ and the axial symmetry the irrotationality condition gives 
\begin{equation}
\partial_r(v_\phi\rho^{c_s^2})=0
\end{equation}
So we get that $v_\phi \rho^{c_s^2}$ is a constant of the motion. As before we perturb the velocities and density around their background values as following
\begin{eqnarray}\label{pert eqs axi v}
v^r(r,t)=v^r_0(r)+v_1^r(r,t)\\ \label{pert eqs axi vt}
v^t(r,t)=v^t_0(r)+v^t_1(r,t)\\ \label{pert eqs axi vphi}
v^\phi(r,t)=v^\phi_0(r)+v^\phi_1(r,t)\\
\rho(r,t)=\rho_0(r)+\rho_1(r,t)\label{pert eqs axi rho}
\end{eqnarray}
Now  using these in $v_\phi\rho^{c_s^2}=\rm{constant}$, we get the relation between $v^\phi_1$ and $\rho_1$ to be 
\begin{equation}\label{vphi1 axi}
 v^\phi_1=-\frac{c_s^2 v^\phi_0}{\rho_0}\rho_1
 \end{equation} 
 Similarly using equations (\ref{pert eqs axi v}), (\ref{pert eqs axi vt}), (\ref{pert eqs axi vphi}) in Eq. (\ref{normalization axi}) and substituting $v^\phi_1$ using Eq. (\ref{vphi1 axi}) we get
 \begin{equation}\label{vt1 axi}
 v^t_1=\alpha_1 v_1^r+\alpha_2 \rho_1
 \end{equation}
 where \begin{equation}
 \alpha_1=\frac{g_{rr}v^r_0}{g_{tt}v^t_0}, \quad \quad \quad \alpha_2=-\frac{g_{\phi\phi}(v^\phi_0)^2c_s^2}{g_{tt}v^t_0\rho_0}
 \end{equation}
For isothermal flow $\frac{p}{\rho}$ is constant. So for vertical equilibrium linear perturbation of Eq. (\ref{H-theta}) gives 
 \begin{equation}
 \frac{(H_\theta)_1}{(H_\theta)_0}=\frac{c_s^2}{\rho_0}\rho_1
 \end{equation}
 where we have used Eq. (\ref{vphi1 axi}) for $v^\phi_1$. As mentioned in Sec. \ref{Accreting black holes as analogue systems} and Sec. \ref{Sec:Definition-mass-accretion-rate} the local flow thickness $H$ for flow with constant thickness and wedge-shaped conical flow does not depend on the accretion variables such as density, pressure or velocity. Therefore $(H_\theta)_1$ representing the perturbation of $H_\theta$ as result of perturbation of pressure, density or velocity is zero for these two kind of flow geometry. Therefore in general, combining these three types of flow geometries, we can write
 \begin{equation}\label{H1}
 \frac{(H_\theta)_1}{(H_\theta)_0}=\frac{\beta}{\rho_0}\rho_1
 \end{equation}
 where $\beta=0$ for both flow with constant thickness and conical flow and $\beta=c_s^2$ for vertical equilibrium.  Substituting $\rho,v^r,v^t$ in the continuity equation, given by Eq. (\ref{cont eq in axi}),  using (\ref{pert eqs axi rho}), (\ref{pert eqs axi v}), and (\ref{pert eqs axi vt}) respectively and retaining only the terms which are linear in perturbation gives
\begin{eqnarray} \label{pert cont axi}
\fl
 \partial_t\left[\sqrt{-\tilde{g}}\left\{v_0^t (H_\theta)_0\rho_1+\rho_0(H_\theta)_0v^t_1+\rho_0v^t_0(H_\theta)_1\right\}\right]\\ \nonumber
 +\partial_r\left[\sqrt{-\tilde{g}}\left\{v_0^r(H_\theta)_0\rho_1+\rho_0(H_\theta)_0v_1^r+\rho_0v^r_0(H_\theta)_1\right\}\right]=0
 \end{eqnarray}
 
\subsection{Perturbation of velocity potential field}\label{Linear-velo-axi}
From irrotationality condition it was found that $v_\phi\rho^{c_s^2}$ is a constant of motion. Also velocity potential is defined as $v_\mu \rho^{c_s^2}=-\partial_\mu \psi$. Therefore $\partial_\phi \psi_1=0$. Thus Eq. (\ref{velocity rho1}) for axially symmetric flow gives
\begin{equation} \label{velo rho1 ax}
\rho_1=\frac{1}{\rho_0^{c_s^2-1}c_s^2}[v_0^t \partial_t\psi_1+v_0^r\partial_r\psi_1]
\end{equation}
and Eq. (\ref{velocity v1}) gives 
\begin{eqnarray}\label{velo vt1 ax}
v_1^t=\frac{1}{\rho_0^{c_s^2}}[(g^{tt}-(v_0^t)^2)\partial_t\psi_1-v^r_0v_0^t\partial_r\psi_1]\\
v_1^r=\frac{1}{\rho_0^{c_s^2}}[(-g^{rr}-(v^r_0)^2)\partial_r\psi_1-v^r_0v_0^t\partial_t\psi_1] \label{velo v1 ax}
\end{eqnarray}
Substituting $\rho_1,v_1^t,v_1^r$ and $(H_\theta)_1$ in Eq. (\ref{pert cont axi}) using equations (\ref{velo rho1 ax}), (\ref{velo vt1 ax}), (\ref{velo v1 ax}) and (\ref{H1}) respectively gives

\begin{eqnarray}
\fl
\partial_t[\tilde{k}_1(r)(-g^{tt}+(1-\frac{1+\beta}{c_s^2})(v_0^t)^2)\partial_t\psi_1]+\partial_t[\tilde{k}_1(r)(1-\frac{1+\beta}{c_s^2})v^r_0v_0^t\partial_r\psi_1] \\ \nonumber
+\partial_r[\tilde{k}_1(r)(1-\frac{1+\beta}{c_s^2})v^r_0v_0^t\partial_t\psi_1]+\partial_r[\tilde{k}_1(r)(g^{rr}+(1-\frac{1+\beta}{c_s^2})(v^r_0)^2)\partial_r\psi_1]=0
\end{eqnarray}
where $\tilde{k}_1(r)=\sqrt{-\tilde{g}}/\rho_0^{c_s^2-1}$. The above equation can be written as $\partial_\mu (f^{\mu\nu}\partial_\nu \psi_1)=0$. Thus $f^{\mu\nu}$ is given by
\begin{equation}
f^{\mu\nu}=\tilde{k}_1\left[\begin{array}{cc}
-g^{tt}+(1-\frac{1+\beta}{c_s^2})(v_0^t)^2 & (1-\frac{1+\beta}{c_s^2})v^r_0v^t_0 \\
(1-\frac{1+\beta}{c_s^2})v^r_0v^t_0 & g^{rr}+(1-\frac{1+\beta}{c_s^2})(v^r_0)^2
\end{array}\right]
\end{equation}

\subsection{Perturbation of the relativistic Bernoulli's constant }\label{Linear-bern-axi}
We defined the variable  $\xi(r,t)$ in Eq. ({\ref{Define-xi}) as $\xi(r,t)=\xi_0+\xi_1(r,t)$. Now using this equation as well as Eq. (\ref{pert eqs axi v}), (\ref{pert eqs axi vt}), (\ref{pert eqs axi vphi}) and (\ref{pert eqs axi rho}) to substitute $\xi,v^r,v^t,v^\phi,\rho$ respectively in the Eq. (\ref{time com euler axi}) and retaining only the terms which are linear in perturbation gives \begin{equation}\label{drxi1 ber axi}
-\frac{\alpha_1 \xi_0}{v_0^r}\partial_t v_1^r-\frac{c_s^2}{\rho_0}\alpha_1 \xi_0 \partial_t \rho_1=\partial_r \xi_1
\end{equation}
Again \begin{equation}\label{xi1 ber axi}
\xi_1=\frac{\xi_0}{v^t_0}v^t_1+\frac{\xi_0c_s^2}{\rho_0}\rho_1.
\end{equation}
Differentiating above equation with respect to $t$ and substituting $v^t_1$ using Eq. (\ref{vt1 axi}) we get
\begin{equation}\label{dtxi ber axi}
\partial_t\xi_1=\frac{\alpha_1 \xi_0}{v^t_0}\partial_t v_1^r+( \frac{\xi_0 c_s^2}{\rho_0}+\frac{\xi_0}{v^t_0}\alpha_2)\partial_t\rho_1
\end{equation} 
 From Eq. (\ref{dtxi ber axi}) and (\ref{drxi1 ber axi}) $\partial_t v_1^r$ and $\partial_t \rho_1$ can be written solely in terms of $\xi_1$
\begin{equation}\label{dtv1 ber axi}
 \partial_t v_1^r=\frac{1}{\Delta}[( \frac{\xi_0 c_s^2}{\rho_0}+\frac{\xi_0}{v^t_0}\alpha_2)\partial_r \xi_1+\frac{\xi_0 \alpha_1c_s^2}{\rho_0}\partial_t\xi_1]
 \end{equation} 
 and 
 \begin{equation}\label{dtrho1 ber axi}
 \partial_t \rho_1=-\frac{\xi_0\alpha_1}{\Delta}[\frac{1}{v_0^t}\partial_r
 \xi_1+\frac{1}{v_0^r}\partial_t\xi_1]
 \end{equation}
 where $\Delta=(\xi_0^2c_s^2\alpha_1)/(v^r_0\rho_0g_{tt}(v_0^t)^2)$. Differentiating the Eq. (\ref{pert cont axi}) with respect to $t$ and rearranging gives
 \begin{eqnarray}
 \fl
  \partial_t\left[\sqrt{-\tilde{g}}(H_\theta)_0\left\{v^t_0 (1+\beta)+\alpha_2 \rho_0\right\}\partial_t\rho_1+\sqrt{-\tilde{g}}(H_\theta)_0\rho_0\alpha_1 \partial_t v_1^r\right]\\ \nonumber
  +\partial_r\left[\sqrt{-\tilde{g}}(H_\theta)_0\left\{v^r_0(1+\beta)\partial_t\rho_1+\rho_0\partial_tv_1^r\right\}\right]=0.
  \end{eqnarray} 
Now substituting $\partial_tv_1^r$ and $\partial_t\rho_1$ in the above equation using (\ref{dtv1 ber axi}) and (\ref{dtrho1 ber axi})  we get
\begin{eqnarray}
\fl
\partial_t[\tilde{k}_2(r)\{-g^{tt}+(1-\frac{1+\beta}{c_s^2})(v_0^t)^2\}\partial_t\xi_1]+\partial_t[\tilde{k}_2(r)v^r_0v^t_0(1-\frac{1+\beta}{c_s^2})\partial_r\xi_1]\\ \nonumber
+\partial_r[\tilde{k}_2(r)v^r_0v^t_0(1-\frac{1+\beta}{c_s^2})\partial_t\xi_1] \\ \nonumber
+\partial_r[\tilde{k}_2(r)\{g^{rr}+(1-\frac{1+\beta}{c_s^2})(v^r_0)^2\}\partial_r\xi_1]=0 
\end{eqnarray}
where $\tilde{k}_2(r)=(\sqrt{-\tilde{g}}(H_\theta)_0)/(\rho_0^{c_s^2-1})$. The above equation is of the form $\partial_\mu \{f^{\mu\nu}\partial_\nu \xi_1\}=0$. So $f^{\mu\nu}$ is found to be
\begin{equation}
f^{\mu\nu}=\tilde{k}_2(r)\left[\begin{array}{cc}
-g^{tt}+\left(1-\frac{1+\beta}{c_s^2}\right)(v_0^t)^2 & v^r_0v^t_0\left(1-\frac{1+\beta}{c_s^2}\right)\\
v^r_0v^t_0\left(1-\frac{1+\beta}{c_s^2}\right) & g^{rr}+\left(1-\frac{1+\beta}{c_s^2}\right)(v^r_0)^2
\end{array}\right]
\end{equation}
\subsection{Perturbation of the Mass Accretion Rate}\label{Linear-mass-axi}
Now let us find the $f^{\mu\nu}$ by perturbing the mass accretion rate. Using $\xi_1$ from Eq. (\ref{xi1 ber axi}) in the Eq. (\ref{drxi1 ber axi}) we have 
\begin{eqnarray}
\frac{\alpha_1}{v_0^r}\partial_tv_1^r+\frac{c_s^2}{\rho_0}\alpha_1 \partial_t\rho_1+\partial_r(\frac{v^t_1}{v_0^t}+\frac{c_s^2}{\rho_0}\rho_1 )=0
\end{eqnarray}
where $v^t_1$ is given by Eq. (\ref{vt1 axi}). Differentiating the above equation with respect to time gives 
\begin{equation}\label{per mass axi}
\partial_t(\frac{\alpha_1}{v_0^r}\partial_tv_1^r )+\partial_t(\frac{\alpha_1 c_s^2}{\rho_0}\partial_t\rho_1 )+\partial_r(\frac{\alpha_1}{v_0^t}\partial_t v_1^r )+\partial_r\{(\frac{\alpha_2}{v_0^t}+\frac{c_s^2}{\rho_0} )\partial_t\rho_1\}=0.
\end{equation}
In Eq. (\ref{Define-mass}) we defined $\Psi(r,t)=\sqrt{-\tilde{g}}v^r(r,t)\rho(r,t)H_\theta(r,t)$ which has the stationary part equal to the mass accretion rate $\Psi_0$ as $ \Psi(r,t)=\Psi_0+\Psi_1(r,t)$. So 
\begin{equation}
 \Psi_1(r,t)=\sqrt{-\tilde{g}}\{\rho_0 v_0^r(H_\theta)_1+\rho_1 v_0^r(H_\theta)_0+\rho_0 v_1^r (H_\theta)_0\}.
 \end{equation} 
Taking derivative of the above equation with respect to $t$ and using Eq. (\ref{H1}) we have 
\begin{equation}
 \frac{\partial_t\Psi_1}{\Psi_0}=(1+\beta) \frac{\partial_t \rho_1}{\rho_0}+\frac{\partial_t v_1^r}{v^r_0}
 \end{equation}
Now substituting the velocities, density  in the continuity Eq. (\ref{cont eq in axi}) using  Eq. (\ref{pert eqs axi v}), (\ref{pert eqs axi vt}), (\ref{pert eqs axi vphi}) and retaining only the terms that are linear in perturbation terms  we have
\begin{equation}
\frac{\partial_r \Psi_1}{\Psi_0}=-[ \{\frac{v_0^t}{v^r_0\rho_0}(1+\beta)+\frac{\alpha_2}{v^r_0} \}\partial_t\rho_1+ \frac{\alpha_1}{v^r_0}\partial_t v_1^r]
\end{equation}
So we can write $\partial_t v_1^r$ and $\partial_t \rho_1$ solely in terms of $\Psi_1$ from above two equations
\begin{equation}
\frac{\partial_t v_1^r}{v^r_0}=\frac{1}{\Lambda}[\{g_{tt}(v_0^t)^2(1+\beta)-g_{\phi\phi}(v_0^\phi)^2c_s^2\}\frac{\partial_t\Psi_1}{\Psi_0}+(1+\beta)g_{tt}v^r_0v^t_0 \frac{\partial_r\Psi_1}{\Psi_0}]
\end{equation}
and 
\begin{equation}
\frac{\partial_t \rho_1}{\rho_0}=-\frac{1}{\Lambda}[ g_{rr}(v^r_0)^2\frac{\partial_t\Psi_1}{\Psi_0}+g_{tt}v^r_0v^t_0\frac{\partial_r\Psi_1}{\Psi_0}]
\end{equation}
where $\Lambda=(1+\beta)+(1+\beta-c_s^2)g_{\phi\phi}(v_0^\phi)^2$.
Substituting $\partial_t v_1^r$ and $\partial_t\rho_1$ in the Eq. (\ref{per mass axi}) using the above two equations we have 
\begin{eqnarray}
\fl
\partial_t[\tilde{k}_3(r)\{-g^{tt}+(1-\frac{1+\beta}{c_s^2})(v_0^t)^2\}\partial_t\xi_1]+\partial_t[\tilde{k}_3(r)v^r_0v^t_0(1-\frac{1+\beta}{c_s^2})\partial_r\xi_1]\\ \nonumber
+\partial_r[\tilde{k}_3(r)v^r_0v^t_0(1-\frac{1+\beta}{c_s^2})\partial_t\xi_1] \\ \nonumber
+\partial_r[\tilde{k}_3(r)\{g^{rr}+(1-\frac{1+\beta}{c_s^2})(v^r_0)^2\}\partial_r\xi_1]=0
\end{eqnarray}
where $\tilde{k}_3(r)=(g_{rr}v_0c_s^2)/(v_0^t\Lambda)$.
The above equation is of the form $\partial_\mu \{f^{\mu\nu}\partial_\nu \Psi_1\}=0$. So $f^{\mu\nu}$ is found to be
\begin{equation}
f^{\mu\nu}=\tilde{k}_3(r)\left[\begin{array}{cc}
-g^{tt}+\left(1-\frac{1+\beta}{c_s^2}\right)(v_0^t)^2 & v^r_0v^t_0\left(1-\frac{1+\beta}{c_s^2}\right)\\
v^r_0v^t_0\left(1-\frac{1+\beta}{c_s^2}\right) & g^{rr}+\left(1-\frac{1+\beta}{c_s^2}\right)(v^r_0)^2
\end{array}\right]
\end{equation}
 
 \section{Augmenting (1+1) metric to (1+3) by two flat dimensions}\label{Appendix:metric-augmentation}
 It was shown that when no symmetry of the background spacetime or the accretion flow was assumed, the linear perturbation of velocity potential leads to the emergence of (1+3) dimensional acoustic metric. However when we consider the symmetries of the given configuration, the acoustic metric obtained through the same perturbation analysis reduces to (1+1) dimension. This indicates that the problem of having (1+1) metric here is not fundamental and this can be removed easily. 
 
Let us for example consider the linear perturbation analysis of mass accretion rate for Michel flow in section (\ref{Linear-pert-mass_michel}). Linear perturbation analysis gives the Eq. (\ref{mass_michel_f_eq}) which we compare to the form $\partial_\mu (f^{\mu\nu}\partial_\nu \Psi_1)=0$ and get the (1+1) dimensional metric $f^{\mu\nu}$. For example, in Eq. (\ref{mass_michel_f_eq}), the term  $\partial_\phi (f^{\phi\phi}\partial_\phi \Psi_1)$ is missing and thus we take $f^{\phi\phi}=0$ but we could also take $f^{\phi\phi}=f^{\phi\phi}(r)$ which does not affect the equation  (\ref{mass_michel_f_eq}). Similarly we could take $f^{\theta\theta}=f^{\theta\theta}(r)$ to obtain the same outcome. Thus we obtain (1+3) dimensional $f^{\mu\nu}$ by taking $f^{\theta\theta}=f^{\theta\theta}(r)$ and $f^{\phi\phi}=f^{\phi\phi}(r)$ keeping other terms unchanged. The ($t-r$) part of the acoustic metric obtained in this way will be given by 
 \begin{equation}
 G_{\mu\nu}=f(r)\left[\begin{array}{cc}
\frac{-c_s^2+g_{rr}(v^r_0)^2(1-c_s^2)}{g_{rr}} & -v^r_0v^t_0(1-c_s^2)\\
-v^r_0v^t_0(1-c_s^2) &  \frac{c_s^2+g_{tt}(v_0^t)^2(1-c_s^2)}{g_{tt}}
\end{array}\right]
\end{equation}
The conformal factor $f(r)$ does not affect the location of acoustic event horizon, causal structure of the spacetime or the acoustic surface gravity. Therefore we can simply use only the relevant matrix part ignoring this factor.
 
\section*{References}
\bibliography{reference_arif}

\end{document}